\documentclass[tighten]{aastex63} % Use linenumbers before submission. Use trackchanges to enable editing commands.
\turnoffedit

\usepackage[varg]{txfonts}
\usepackage{multirow}
\usepackage{tabularx} % Is said to be necessary to specify both the column width and the horizontal alignment of the text.
\usepackage{capt-of}

\newcolumntype{E}[1]{>{\raggedright\arraybackslash}p{#1}} % With this definition, the command E{2cm} makes a column of width 2cm whose text is aligned left. Is said to need tabularx. Originally, it was L instead of E. But in ApJ, L is another command.
\newcolumntype{N}[1]{>{\centering\arraybackslash}p{#1}} % With this definition, the command N{2cm} makes a column of width 2cm whose text is centred. Is said to need tabularx. Originally, it was C instead of N. But in ApJ, C is another command.
\newcolumntype{I}[1]{>{\raggedleft\arraybackslash}p{#1}} % With this definition, the command I{2cm} makes a column of width 2cm whose text is aligned right. Is said to need tabularx. Originally, it was R instead of I. But in ApJ, R is another command.

\usepackage{natbib,twoopt}
\bibpunct{(}{)}{;}{a}{}{,}             %% natbib format for A&A and ApJ
\makeatletter
  \newcommandtwoopt{\citeads}[3][][]{\href{http://adsabs.harvard.edu/abs/#3}%
    {\def\hyper@linkstart##1##2{}%
     \let\hyper@linkend\@empty\citealp[#1][#2]{#3}}}
  \newcommandtwoopt{\citepads}[3][][]{\href{http://adsabs.harvard.edu/abs/#3}%
    {\def\hyper@linkstart##1##2{}%
     \let\hyper@linkend\@empty\citep[#1][#2]{#3}}}
  \newcommandtwoopt{\citetads}[3][][]{\href{http://adsabs.harvard.edu/abs/#3}%
    {\def\hyper@linkstart##1##2{}%
     \let\hyper@linkend\@empty\citet[#1][#2]{#3}}}
  \newcommandtwoopt{\citeyearads}[3][][]%
    {\href{http://adsabs.harvard.edu/abs/#3}
    {\def\hyper@linkstart##1##2{}%
     \let\hyper@linkend\@empty\citeyear[#1][#2]{#3}}}
\makeatother
\received{2021 April 22}
\revised{2021 June 4}
\accepted{2021 June 9}
\published{2021}
\submitjournal{ApJ}

\shorttitle{Pair Cascades at the Edge of 
the Broad-Line Region Shaping the $\gamma$-ray Spectrum of 3C\,279}

\shortauthors{Wendel, Shukla and Mannheim}

\begin{document}

\title{Pair Cascades at the Edge of the Broad-
%emission 
Line Region Shaping the Gamma-Ray Spectrum of 3C\,279}

\author[0000-0003-2898-2731]{Christoph Wendel}
\affiliation{Institut für Theoretische Physik und Astrophysik, Universit\"at W\"urzburg, Emil-Fischer-Stra{\ss}e 31, D-97074 W\"urzburg, Germany}

\correspondingauthor{Christoph Wendel}
\email{cwendel@astro.uni-wuerzburg.de}

\author[0000-0002-5656-2657]{Amit Shukla}
\affiliation{Discipline of Astronomy, Astrophysics and Space Engineering, Indian Institute of Technology Indore, Khandwa Road, Simrol, Indore, India 453552}

\author{Karl Mannheim}
\affiliation{Institut für Theoretische Physik und Astrophysik, Universit\"at W\"urzburg, Emil-Fischer-Stra{\ss}e 31, D-97074 W\"urzburg, Germany}

\begin{abstract}
% Context:
The blazar 3C\,279 emits a flux of gamma-rays that is variable on timescales as short as the light-crossing time across the event horizon of its central black hole. It is commonly reported that the spectral energy distribution (SED) does not show signs of pair attenuation due to interactions of the gamma rays with ambient ultraviolet photons, concluding that the gamma rays must originate from substructures in the jet outside of the broad-line region (BLR).
% Aims:
We address the spectral signature imprinted by atomic emission lines on the gamma-ray spectrum produced by an inverse-Compton pair cascade in the photon field of the BLR.
% Methods:
We determine with high precision the gamma-ray SED of 3C\,279 using Fermi Large Area Telescope data from MJD 58129--58150 and simulate the pair cascade spectrum for three different injection terms.
%Three different hard-spectrum injection terms are used in the simulation.
% Results:
Satisfactory fits to the observational data are obtained. The obtained SED shows features imprinted by pair production on atomic emission line photons due to optically thick radiation transport, but lacking further exponential attenuation expected if the emission region would lie buried deep within the BLR.  
% Conclusion:
The SED of 3C\,279 is consistent with an inverse-Compton pair cascade spectrum without exponential external pair absorption. Our findings support the view that the gamma-ray emission in 3C~279 originates from the edge of the BLR.
\end{abstract}


\section{Introduction} \label{SectionIntroduction}

Since the first detection of gamma-ray emission from the flat-spectrum radio quasar (FSRQ) 3C\,279 in 1991 by the energetic gamma ray experiment telescope (EGRET) \citepads{1992ApJ...385L...1H}, the location and extent of its originating region is debated. Recently, the inner jet of 3C\,279 was imaged at millimeter wavelengths with an unprecedented angular resolution of $\approx 20 \, \mu \rm as$ \citepads{2020A&A...640A..69K}. The resolution in the gamma-ray regime is, however, orders of magnitude bigger than the resolution necessary to directly dissect the morphological structure of the active galactic nucleus (AGN). Thus, only indirect arguments can be brought forth in the discussion on the nature of the emission region of gamma rays from blazars. 

\added{It is well known that the high photon density in the launching region of the jets leads to a $\gamma$-sphere with a radius that increases with photon energy \citepads{1995ApJ...441...79B}.} The effect of absorption on the observed gamma-ray spectra of quasars due to pair production (PP) in collisions of their GeV photons with photons from the ultraviolet (UV) to X-ray radiation field within the broad emission line region (BLR) has been studied by \citetads{2003APh....18..377D} and by \citetads{2006ApJ...653.1089L}. The latter authors constructed a spherical, shell-like BLR model and took observed emission lines as well as continuum blackbody radiation into account. By determining the pair-absorption (= absorption of photons due to PP) optical depth, they concluded that detection of radiation from $10 \, \rm GeV$ to some $100 \, \rm GeV$ from 3C\,279 (and FSRQs in general) would imply emission from a distance of several $10^{15} \, \rm m$ away from the central object, hence not from inside the BLR. If radiation above $10 \, \rm{GeV}$ was emitted inside the BLR, it would be completely absorbed. 

A similar approach was pursued by \citetads{2007ApJ...665.1023R} using a double-absorber model and including redshift dependence, by \citetads{2009MNRAS.392L..40T} and \citetads{2012arXiv1209.2291T}, who used the photoionization code Cloudy, and by \citetads{2015arXiv150207624B}, who used the six strongest UV lines, to model BLR spectra of 3C\,279 and optical depths, which are found to be $\approx 10$ above few tens of $\rm GeV$ and for emission from within the BLR. Less severe limits were found by \citetads{2016ApJ...821..102B}, who determined the optical depth of a spherical, shell-like BLR of 3C\,279 to be $> 1$ only inside the inner boundary and only for energies $> 50 \, \rm GeV$.
%Follow-up studies by several authors also found the BLR to be optically thick for gamma rays above few $10 \, \rm GeV$.

In their extensive variability study of the brightest flares of 3C\,279 and other Fermi-detected FSRQs, \citetads{2019ApJ...877...39M} found no significant imprint of absorption, and determine the minimum distance of the emission region to several $10^{14} \, \rm m$, in agreement with estimates based on variability and on the comparison of cooling times with flare decays. Lack of absorption was also reported for a Fermi- and high-energy (HE) stereoscopic system (H.E.S.S.)-detected 2015 June flare, placing the emitting region over $10^{15} \, \rm m$ \citepads{2019A&A...627A.159H}, as well as for the flare-in-flare producing magnetic reconnection events on timescales of minutes from 2018 (\citeads{2020NatCo..11.4176S}; \citeads{2021ApJ...912...40M}).

By fitting a one-zone, synchrotron-self-Compton (SSC) plus external-Compton (EC) model to quasi-simultaneous FSRQ spectral energy distributions (SEDs), \citetads{2020ApJS..248...27T} inferred a distance of some $10^{16} \, \rm m$ for 3C\,279. A pure EC model was used by \citetads{2019MNRAS.484.3168S} to fit a 2018 flare, concluding that scattering occurs on photons from the dusty torus.

Evidence for radiation production at the outer BLR edge of 3C 279 was found by \citetads{2014ApJ...782...82D} through fitting quasi-simultaneous SEDs from 2008--2009 with a leptonic model under the assumptions of a log-parabolic electron distribution and of equipartition between nonthermal leptons and magnetic field energy density.

Indications for emission of gamma rays from inside the BLR were found by \citetads{2010ApJ...717L.118P}, who could fit Fermi-detected FSRQs including 3C\,279 with a double-absorbed power law (PL) and associate the GeV breaks to pair absorption on H I Lyman continuum and He II Lyman continuum photons. Both a 2013 flare registered by the Fermi large area telescope (LAT) as well as a 2015 flare detected by the Astrorivelatore Gamma a Immagini Leggero (AGILE) from 3C\,279 could be described by a one-zone SSC + EC model applying emission within the BLR (\citeads{2015ApJ...807...79H}; \citeads{2018ApJ...856...99P}). The HE emission of a 2011 high-activity state was ascribed by \citeads{2014A&A...567A..41A} to EC emission mainly on BLR photons. \citetads{2021MNRAS.500.5297A} found threefold evidence for inside-BLR gamma-ray emission of the June 2015 flare of 3C 279, namely, by hour-scale variability and assuming an emission region size of the jet cross section \citepads[see also][]{2016ApJ...824L..20A}, by the preference of a log-parabolic spectrum over a PL, and by achromatic cooling. The highest energetic photon energy of $\approx 89 \, \rm{GeV}$ detected in their study is also compatible with the BLR being the origin of gamma rays above $20 \, \rm{GeV}$.

In all the above-named absorption studies, the escaping spectrum and the question whether it is detectable or not is dependent on the shape of the injected (intrinsic) spectrum, which in most cases is assumed
%ad hoc
to be a PL,
%of certain index
a broken PL, or a log-parabola. Furthermore, in these works only pair absorption is considered. The induced cascade and especially the inverse-Compton (IC) up-scattering of photons by the pair-produced electrons is not taken into account in the above-mentioned absorption studies. In their jet models, \citetads{1995MNRAS.277..681M} and \citetads{2018A&A...620A..41V} have stressed the necessity to consider both a radiation transport term $\sim (1 - \exp(-\tau))/\tau$ for the emerging flux from the jet with a constant source function inside of the jet and an additional term $\sim \exp(-\tau)$ to account for the absorption by the jet and external radiation field outside of the jet.

In section \ref{SectionModeling}, we study the escaping HE spectra of 3C\,279 from IC pair cascades considering three different cases of the injected particle distributions.
%We inject PL photons plus Gaussian electrons in case 1, logparabolic photons in case 2, and PL plus Gaussian electrons in case 3.
The IC pair cascade equations are numerically solved assuming BLR photons as the main source of soft target photons. We consider not only pair absorption, but also reprocessed emission by the cascade in the BLR field, in contrast to pure absorption by an external screen $\sim \exp(-\tau)$. In section \ref{SectionResults}, we compare the simulated spectra from our numerical code with an observed 
%high-fidelity 
spectrum of
%the FSRQ
3C~279 obtained from the analysis of Fermi LAT data in section \ref{SectionDataAnalysis}.  
%and find that the IC up-scattered component is non-negligible for the escaping spectra.
In section \ref{SectionSummary}, we summarize our findings.

In what follows we denote by $m_{\rm e}$, $c$, and $h$ the electron rest mass, the velocity of light, and Planck's constant, respectively. We use the redshift $z = 0.54$ of 3C\,279 \citepads{1996ApJS..104...37M} and the luminosity distance $D = 3.1 \, \rm{Gpc}$.

\section{Data Analysis} \label{SectionDataAnalysis}

The LAT \citepads{2009ApJ...697.1071A} on board the Fermi spacecraft is a pair-conversion gamma-ray telescope. The LAT energy sensitivity ranges from 20 MeV to 300 GeV with a 2.5 sr large field of view. We have analyzed the pass8 Fermi LAT gamma-ray data of the source using (Science Tools version v10r0p5). A \textit{region of interest} with a circular radius of $10^\circ$ around 3C~279 was selected for analysis. A zenith angle cut of $90^\circ$, the GTMKTIME cut of DATA\_QUAL==1 $\&\&$ LAT\_CONFIG==1 together with the LAT event class ==128 and the LAT event type ==3 was used. The spectral analysis on the resulting data set was carried out by including {\it gll\_iem\_v06.fits} and the isotropic diffuse model {\it iso\_P8R2\_SOURCE\_V6\_v06.txt}. The flux and spectrum of 3C~279 were determined by fitting a log-parabola model, using an unbinned gtlike algorithm based on the NewMinuit optimizer (\citeads{1979ApJ...228..939C}; \citeads{1996ApJ...461..396M}). The Fermi LAT data between MJD 58129 and MJD 58150 during a flaring state were analyzed and the SED was generated\added{ in the AGN frame, see blue crosses in figure \ref{FigureFluxDensity}}.

%%%%%%%%%%%%%% Figure for ApJ (Begin) %%%%%%%%%%%%%%
%\begin{figure}
%%\resizebox{\hsize}{!}{\includegraphics[angle=-90]{C:/Chris's/PhyMaLaChePyC/Physik/Materialien, Texte/2020 - ApJ Letter/FigureGeometrySketch3al.pdf}}
%%\resizebox{\hsize}{!}{\includegraphics[angle=-90]{C:/Chris's/PhyMaLaChePyC/Physik/Materialien, Texte/2020 - ApJ Letter/FigureGeometrySketch3bl.pdf}}
%\resizebox{\hsize}{!}{\includegraphics[angle=-90]{FigureGeometrySketch3al.pdf}}
%\resizebox{\hsize}{!}{\includegraphics[angle=-90]{FigureGeometrySketch3bl.pdf}}
%\caption{\added{Sketch of gamma-ray emission models. Upper panel: setting for cases 1\textsubscript{a}, 2, and 3. Lower panel: setting for cases 1\textsubscript{b} and 1\textsubscript{c}.}}
%\label{FigureSketch}
%\end{figure}
%%%%%%%%%%%%%% Figure for ApJ (End) %%%%%%%%%%%%%%%%

%%%%%%%%%%%%%% Figure for arXiv (Begin) %%%%%%%%%%%%%%
\begin{figure*}
\parbox{.41\linewidth}{
\includegraphics[angle=-90,width=0.41\textwidth]{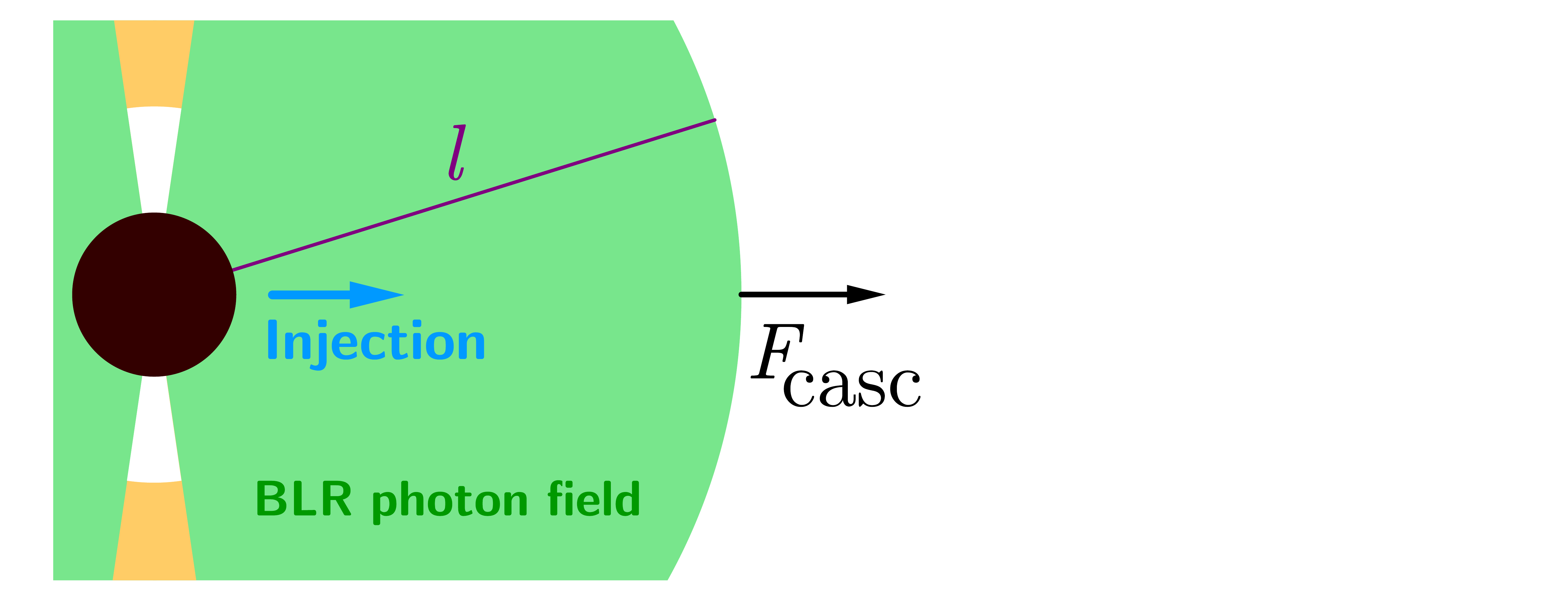}
\vskip0.8cm
\includegraphics[angle=-90,width=0.41\textwidth]{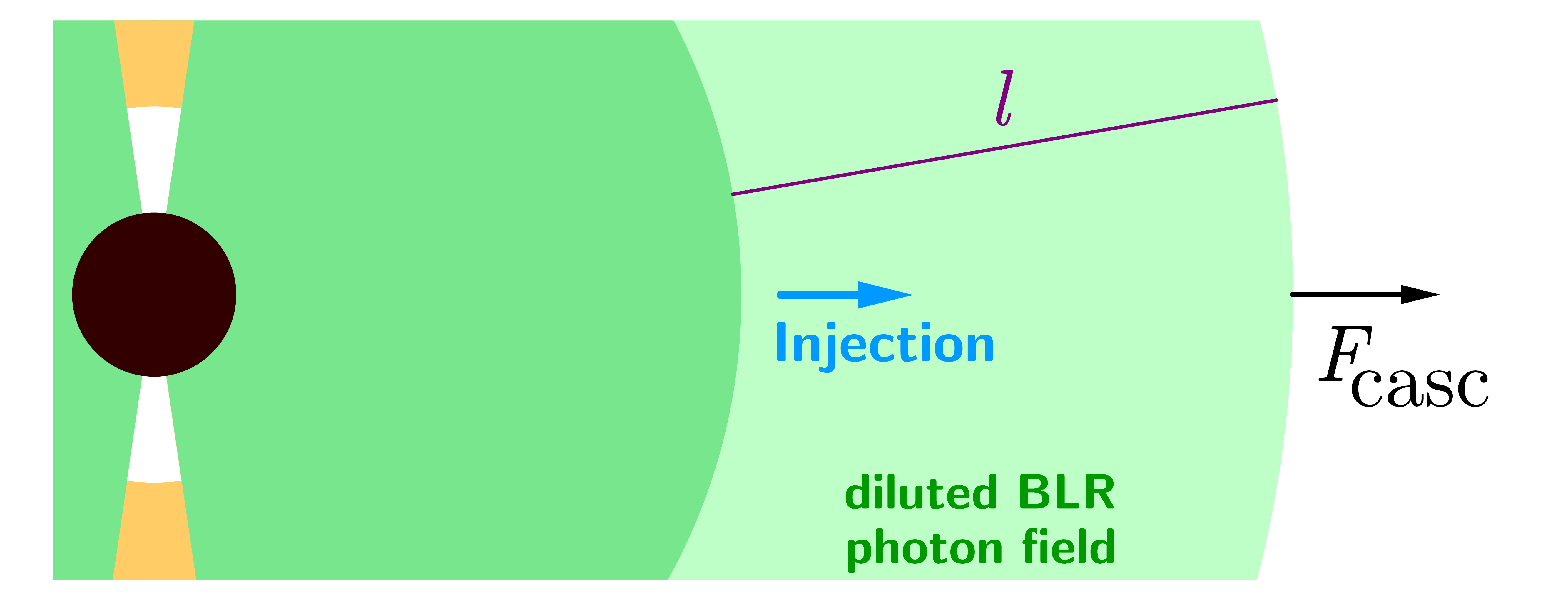}
\vskip0.3cm
\caption{\added{Sketch of gamma-ray emission models.\\Upper panel: setting for cases 1\textsubscript{a}, 2, and 3.\\Lower panel: setting for cases 1\textsubscript{b} and 1\textsubscript{c}.}}
\label{FigureSketch}
}
\hfill
\parbox{.58\linewidth}{
\includegraphics[width=0.58\textwidth]{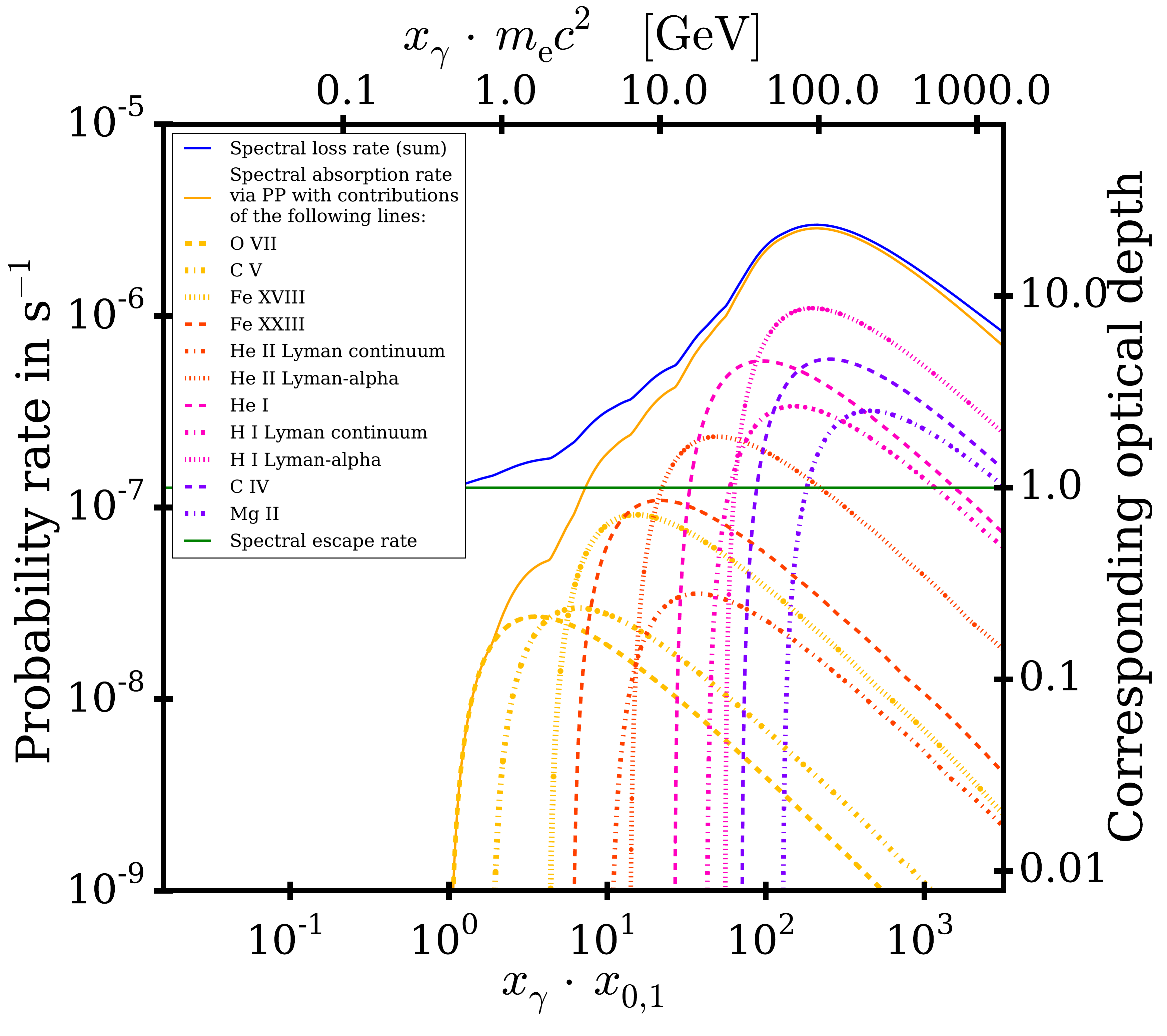}
\caption{Various channels of the photon loss rate for case \replaced{1a}{1\textsubscript{a}} on the left-hand side ordinate vs. the HE photon energy. On the right-hand side ordinate, we show the optical depth corresponding to the respective photon loss rate contribution.}
\label{FigureLossRate}
}
\end{figure*}
%%%%%%%%%%%%%% Figure for arXiv (End) %%%%%%%%%%%%%%%%

\section{Modeling of Inverse-Compton Pair Cascades in the 3C~279 Broad-Line Region} \label{SectionModeling}

We assume that relativistic pairs or gamma rays are injected into the jet plasma of 3C\,279 initiating IC pair cascades on the soft photons of the BLR\added{, see upper panel of figure \ref{FigureSketch}}.
%%%%%%%%% The following sentences were removed: %%%%%%%%%
%The injection of non-thermal energy can result from a variety of processes (cf. e.g. \citeads{1995A&A...295..613M}; \citeads{2002A&A...396..833R}; \citeads{2006A&A...450..887G}; \citeads{2014APh....56....9E}; \citeads{2017SSRv..207..291B}; \citeads{2020NewAR..8901543M}), including magnetospheric lepton acceleration in a sporadically active vacuum gap \citepads[][and references therein]{2021ApJ...908...88H}.
%%%%%%%%% The following sentences were substituted by the preceding sentence to motivate non-thermal distributions: %%%%%%%%%
The injection of nonthermal energy can result from processes such as diffusive shock acceleration (\citeads{2012ApJ...745...63S}; \citeads{2014ApJ...780...87M}; \citeads{2017MNRAS.464.4875B}), stochastic acceleration (\citeads{2011ApJ...740...64L}; \citeads{2015ApJ...808L..18A}), magnetoluminescence (\citeads{2017SSRv..207..291B}; \citeads{2020NewAR..8901543M}), shear acceleration (\citeads{2002A&A...396..833R}; \citeads{2007Ap&SS.309..119R}; \citeads{2019Galax...7...78R}), wakefield acceleration by Alfv\'en waves excited at the base of the jet \citepads{2014APh....56....9E}, or by acceleration in kink-unstable magnetically dominated flows (\citeads{2000A&A...355..818A}; \citeads{2006A&A...450..887G}), see the review by \citetads{2019ARA&A..57..467B}. A Gaussian electron distribution can arise from particle acceleration in a sporadically active magnetospheric vacuum gap \citepads[][and references therein]{2021ApJ...908...88H}.
%(\citeads{2018Galax...6..122H}; \citeads{2018A&A...616A.184L}; \citeads{2020ApJ...895..121C}; \citeads{2020ApJ...895...99K})
We note that a proton beam propagating through the central parsec could inject secondaries where in situ acceleration of electrons and positrons to high energies is quenched due to strong energy losses (\citeads{1976Natur.262..649L}; \citeads{1993A&A...269...67M}; \citeads{1995A&A...295..613M}). We inject PL photons plus Gaussian electrons in \replaced{case 1}{cases 1\textsubscript{a}, 1\textsubscript{b} and 1\textsubscript{c}}, log-parabolic photons in case 2, and PL plus Gaussian electrons in case 3.

\replaced{The}{Similar to the approaches by \citetads{1988ApJ...335..786Z} and \citetads{1995ApJ...441...79B}, the} emerging spectra from this scenario can be numerically computed with a code that solves the coupled kinetic equations for the repeated interactions of relativistic electrons (positrons) and gamma-ray photons with a background field of low-energy photons \citepads{2021A&A...646A.115W}. Three homogeneous, isotropic, and time-independent distributions are included: the distribution of relativistic electrons with Lorentz factor $\gamma \gg 1$ and spectral number density $N(\gamma)$, the distribution of gamma-ray photons with energy $x_\gamma \gg 1$ and spectral number density $n_\gamma (x_\gamma)$ and the distribution of low-energy photons with energy $x \ll 1$ and spectral number density $n_0(x)$, which is assumed to be a set of broad emission lines in the following. Energies are in units of $m_{\rm e} c^2$. Pair production involves collisions of the gamma-ray photons with low-energy photons, destroying the incident photons and supplying new electrons. IC scattering involves collisions of electrons with low-energy photons. The electrons lose energy but remain in the system and can IC scatter again. The gamma-ray photons produced are available for another generation of PP. The spectral IC production rate of gamma-ray photons per unit volume is denoted by $\dot n_{\gamma, \, \rm{IC}}(x_\gamma)$. Allowing injection of gamma-ray photons with spectral rate $\dot n_{\gamma , \, \rm i}(x_{\gamma})$ as well as injection of electrons with spectral rate $\dot N_{\rm i}(\gamma)$, an IC pair cascade develops through the interplay of PP and IC scattering, affecting the relativistic electrons and the gamma-ray photons, while the soft target photons have a fixed density. For the escape of the electrons and photons from the interaction region, an energy-independent escape time is adopted. Hence, photons disappear through two channels, namely, by pair absorption and by escape. Instead of assuming that the gamma-ray emitting region is located outside of the BLR, we assume that it lies at the edge of the BLR\added{, as depicted in the upper panel of figure \ref{FigureSketch},} such that the BLR photons act as target photons for the cascade, but do not lead to further pair absorption beyond the emission region.
%The governing kinetic equations as well as the scheme to solve them numerically are laid out in \citetads{2021A&A...646A.115W}. Isotropic, homogeneous and time-independent distributions are presumed.

%%%%%%%%%%%%%% Figure for ApJ (Begin) %%%%%%%%%%%%%%
%\begin{figure}
%\resizebox{\hsize}{!}{\includegraphics{C:/Chris's/PhyMaLaChePyC/Physik/Astro - Modelling 3C279/Photon-loss-rate - Probability-rate versus photon energy from 2020-09-02 12-49 4.pdf}}
%%\resizebox{\hsize}{!}{\includegraphics{Photon-loss-rate - Probability-rate versus photon energy from 2020-09-02 12-49 4.pdf}}
%\caption{Various channels of the photon loss rate for case \replaced{1a}{1\textsubscript{a}} on the left-hand side ordinate vs. the HE photon energy. On the right-hand side ordinate, we show the optical depth corresponding to the respective photon loss rate contribution.}
%\label{FigureLossRate}
%\end{figure}
%%%%%%%%%%%%%% Figure for ApJ (End) %%%%%%%%%%%%%%%%

To study the effect of the many possible acceleration mechanisms on the emerging SED, we assume three different generic injection terms:

Case 1: gamma-ray photons from the inner portion of 3C\,279 as well as electrons that have been accelerated to energies around a mean energy $\gamma_{\rm{mean}}$ with spread $\varsigma$ in energy (e.g. from a voltage drop of a presumed magnetospheric vacuum gap) interact with photons from the BLR. Hence, as input to the code we prescribe $\dot n_{\gamma , \, \rm i}$ and $\dot N_{\rm i}$. For the photon injection, we use a PL
\begin{equation}
\dot n_{\gamma , \, \rm i}(x_{\gamma}) = \left\{
\begin{array}{ll}
K_{{\rm P}} \left( \frac{x_{\gamma}}{x_{\gamma, \, 1}}\right) ^{\alpha} & \mathrm{if} \; x_{\gamma, \, 1} \leq x_{\gamma} \leq x_{\gamma, \, 0} \\
0 & \mathrm{otherwise,}
\end{array}
\right.
\label{EquationHEPInjectionScenario1}
\end{equation}
while we model the electron distribution by a Gaussian
\begin{equation}
\dot N_{\rm i}(\gamma) = \left\{
\begin{array}{ll}
\frac{K_{\rm{G}}}{\varsigma \sqrt(2 \pi)} \cdot \exp \left( -\frac{(\gamma-\gamma_{\mathrm{mean}})^2}{2 \, \varsigma^2} \right) & \mathrm{if} \; \gamma_{{\mathrm{i}},\,1} \leq \gamma \leq \gamma_{{\mathrm{i}},\,0} \mathrm{,} \\
0 & \mathrm{otherwise.}
\end{array}
\right.
\label{EquationElectronInjectionScenario1}
\end{equation}
\added{In the following, we consider the three sub-cases 1\textsubscript{a}, 1\textsubscript{b} and 1\textsubscript{c}. The common property of the sub-cases is that the gamma-ray and electron injection functions are given by equations \ref{EquationHEPInjectionScenario1} and \ref{EquationElectronInjectionScenario1}. Their distinguishing features are different values of the parameters $K_{\rm{lines}}$, $K_{{\rm P}}$, $\alpha$ and $\varsigma$, see section \ref{SectionResults}.}

Case 2: Only gamma-ray photons interact with BLR photons. One has to specify $\dot n_{\gamma , \, \rm i}$, while $\dot N_{\rm i} = 0$. We restrict the photon spectral injection rate to a log-parabola:
\begin{equation}
\dot n_{\gamma , \, \rm i}(x_{\gamma}) = \left\{
\begin{array}{ll}
K_{{\rm P}} \left( \frac{x_{\gamma}}{x_{\gamma, \, 1}}\right) ^{\alpha+\beta \log_{10}(x_{\gamma} / x_{\gamma, \, 1})} & \mathrm{if} \; x_{\gamma, \, 1} \leq x_{\gamma} \leq x_{\gamma, \, 0} \\
0 & \mathrm{otherwise}
\end{array}
\right.
\label{EquationHEPInjectionScenario2}
\end{equation}

Case 3: Purely electrons interact with photons from the BLR. Only $\dot N_{\rm i}$ has to be defined and $\dot n_{\gamma , \, \rm i} = 0$. The electron distribution is prescribed by a PL plus Gaussian:
%%%%%%%%%%%%%% Eq for ApJ (Begin) %%%%%%%%%%%%%%
%\begin{eqnarray}
%& &\dot N_{\rm i}(\gamma) = \label{EquationElectronInjectionScenario3} \\
%& &\left\{
%\begin{array}{ll}
%K_{{\rm P}} \gamma^{\alpha} + \frac{K_{\rm{G}}}{\varsigma \sqrt(2 \pi)} \cdot \exp \left( -\frac{(\gamma-\gamma_{\mathrm{mean}})^2}{2 \, \varsigma^2} \right) & \mathrm{if} \; \gamma_{{\mathrm{i}},\,1} \leq \gamma \leq \gamma_{{\mathrm{i}},\,0} \mathrm{,} \\
%0 & \mathrm{otherwise}
%\end{array}
%\right. \nonumber
%\end{eqnarray}
%%%%%%%%%%%%%% Eq for ApJ (End) %%%%%%%%%%%%%%%%

%%%%%%%%%%%%%% Eq for arXiv (Begin) %%%%%%%%%%%%%%
\begin{equation}
\dot N_{\rm i}(\gamma) = \left\{
\begin{array}{ll}
K_{{\rm P}} \gamma^{\alpha} + \frac{K_{\rm{G}}}{\varsigma \sqrt(2 \pi)} \cdot \exp \left( -\frac{(\gamma-\gamma_{\mathrm{mean}})^2}{2 \, \varsigma^2} \right) & \mathrm{if} \; \gamma_{{\mathrm{i}},\,1} \leq \gamma \leq \gamma_{{\mathrm{i}},\,0} \mathrm{,} \\
0 & \mathrm{otherwise}
\end{array}
\right. \label{EquationElectronInjectionScenario3}
\end{equation}
%%%%%%%%%%%%%% Eq for arXiv (End) %%%%%%%%%%%%%%%%

%In all cases, $n_0$ has to be given.
%Similarly to \citetads{2021A&A...646A.115W}, 
In all\deleted{ three} cases we choose
\begin{equation}
n_0(x) = K_{\rm{lines}} \cdot \sum _i \frac{K_{{\rm{line}},\,i}}{x_{0,\,i}} \cdot \delta_{\mathrm{Dirac}} \left( x - x_{0,\,i} \right)
\label{EquationDistributionLEPs}
\end{equation}
being a sum of Dirac delta distributions situated at mid-UV (MUV) to soft X-ray energies $x_{0,\,i} = h / (\lambda_{0,\,i} \, m_{\rm{e}} \, c)$. This approximates that the background photon field $n_0$ is a set of emission lines at the wavelengths $\lambda_{0,\,i}$. Each emission line is defined by its photon energy $x_{0,\,i}$ and by $K_{{\rm{line}},\,i}$, which is its flux density relative to that of the hydrogen Balmer-$\beta$ line. $K_{\rm{lines}}$ is a parameter affecting the total number density of the soft photons. The fact that the cascade evolves with BLR photons as targets reflects that the cascade happens within or at least not far away from the BLR.

%%%%%%%%%%%%%% Figure for ApJ (Begin) %%%%%%%%%%%%%%
%\begin{figure}
%\resizebox{\hsize}{!}{\includegraphics{C:/Chris's/PhyMaLaChePyC/Physik/Astro - Modelling 3C279/Dotn times xgamma^2 versus final photon energy from 2020-09-02 12-46.pdf}}
%%\resizebox{\hsize}{!}{\includegraphics{Dotn times xgamma^2 versus final photon energy from 2020-09-02 12-46.pdf}}
%\caption{Product of the squared HE photon energy with the spectral production rate of photons (dotted), with the injection rate (dashed), as well as with the sum of both (solid) vs. the product of the HE photon energy with the energy of the highest energetic line (bottom abscissa) and vs. the HE photon energy in GeV (top abscissa). All curves are for \replaced{case1a}{case 1\textsubscript{a}}. For the exact meaning of the plotted quantities, see section \ref{SectionModeling} or \citetads{2021A&A...646A.115W}.}
%\label{FigureHEPDistribution1}
%\end{figure}
%%%%%%%%%%%%%% Figure for ApJ (End) %%%%%%%%%%%%%%%%

%%%%%%%%%%%%%% Figure for arXiv (Begin) %%%%%%%%%%%%%%
\begin{figure*}
\parbox{.49\linewidth}{
\includegraphics[width=0.48\textwidth]{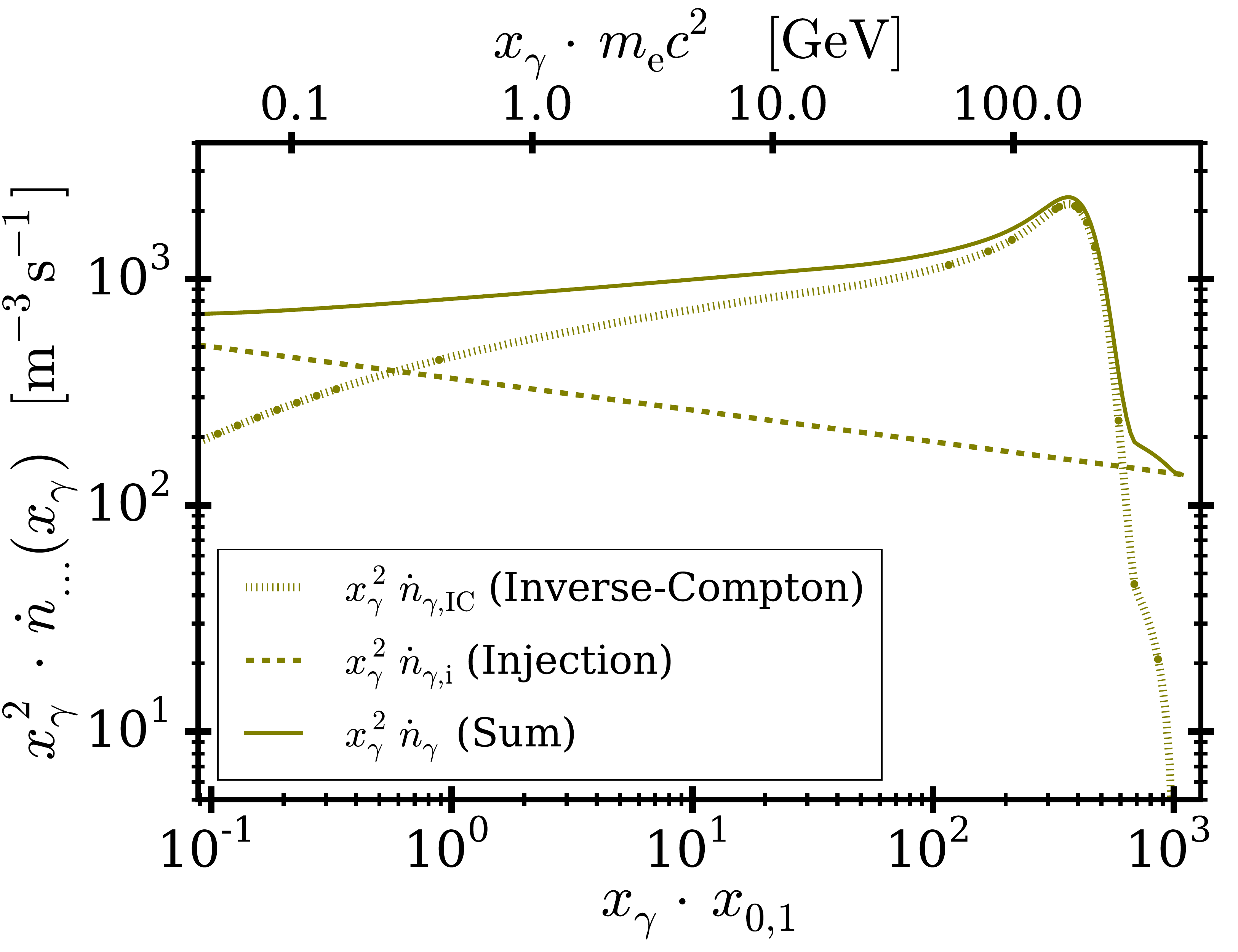}
\caption{Product of the squared HE photon energy with the spectral production rate of photons (dotted), with the injection rate (dashed), as well as with the sum of both (solid) vs. the product of the HE photon energy with the energy of the highest energetic line (bottom abscissa) and vs. the HE photon energy in GeV (top abscissa).}
\label{FigureHEPDistribution1}
}
\hfill
\parbox{.49\linewidth}{
\includegraphics[width=0.48\textwidth]{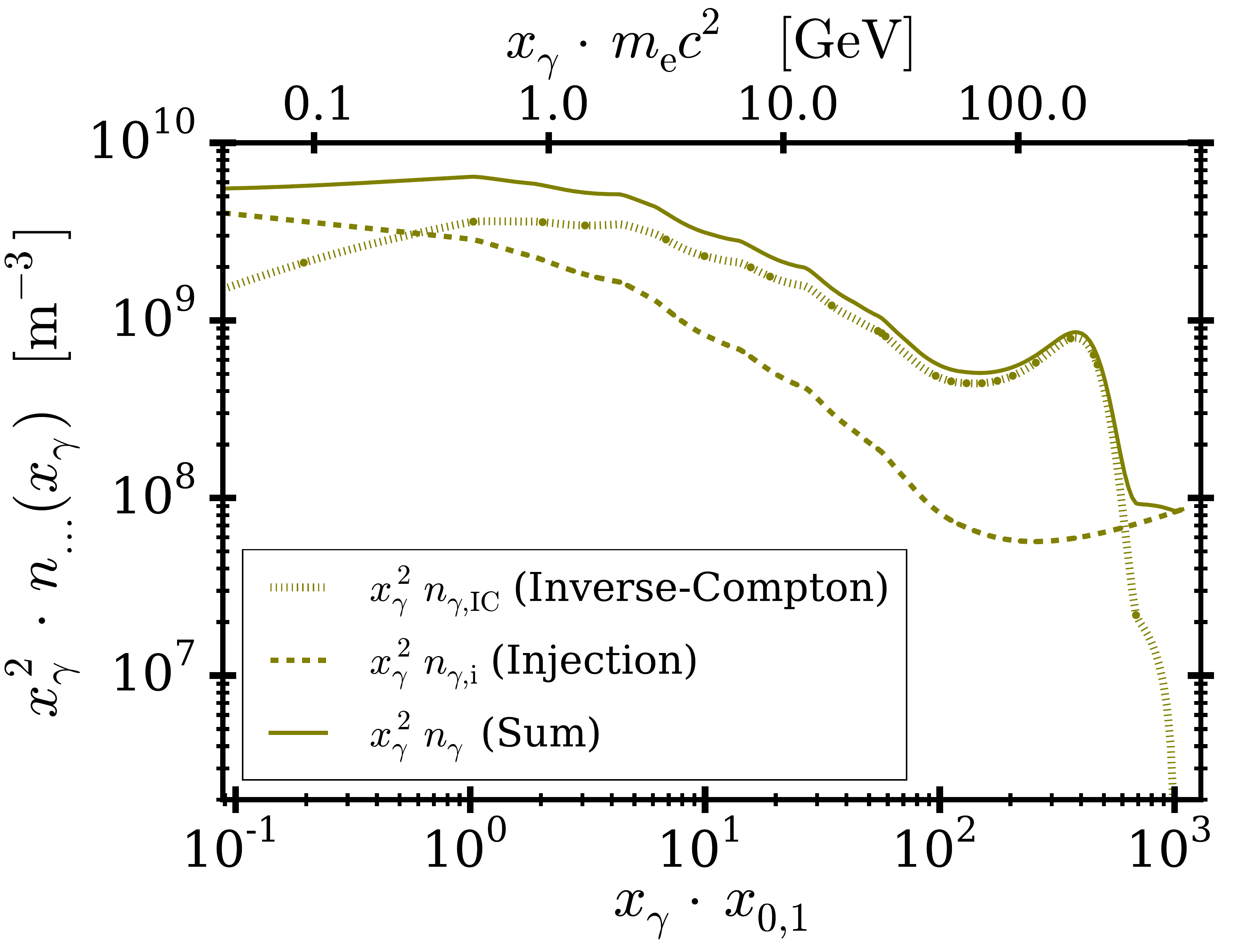}
\caption{Product of the squared HE photon energy with the spectral number density of photons (solid) as well as with its two contributions $n_{\gamma,\,\mathrm{i}}$ (dashed) and $n_{\gamma,\,\mathrm{IC}}$ (dotted) vs. the product of the HE photon energy with the energy of the highest energetic line (bottom abscissa) and vs. the HE photon energy in GeV (top abscissa).}
\label{FigureHEPDistribution2}
}
\begin{center}\vspace{-0.3cm}
All curves are for case \replaced{1a}{1\textsubscript{a}}. For the exact meaning of the plotted quantities, see section \ref{SectionModeling} or \citetads{2021A&A...646A.115W}.
\end{center}
\end{figure*}
%%%%%%%%%%%%%% Figure for arXiv (End) %%%%%%%%%%%%%%%%

%Analogously to Eq. 7 of \citetads{2021A&A...646A.115W}
The escape timescale of both the electrons and the photons is approximated by \replaced{$T_{\rm{esc}} := R / c$, where $R$ is the energy-independent, radial size of the interaction region}{$T_{\rm{esc}} := l / c$, where $l$ is the energy-independent escape length (identified with the radius $R$ of the interaction region in \citetads{2021A&A...646A.115W})}. As we assume the injection taking place in the interior of the BLR, the quantity \replaced{$R$}{$l$} is also equivalent to the radial size of the BLR\added{, see upper panel of figure \ref{FigureSketch}}. Outside of the radial distance \replaced{$R$}{$l$}, the radiation is assumed to escape without additional absorption due to the $r^{-2}$ dilution of the BLR field, see the discussion in \citetads{1995MNRAS.277..681M} and \citetads{2018A&A...620A..41V}.

Eventually, from the injection rates, from $n_0$, and from $T_{\rm{esc}}$, the code iteratively determines the steady-state $N$. We iterate the values of $N$ as long as the relative change of $N$ between two successive iteration steps is $\geq 0.01$ for $\gamma > 10 / x_{0,\,1}$ and $\geq 0.01 \cdot (\gamma \, x_{0,\,1}/10)^2$ for $\gamma \leq 10 / x_{0,\,1}$. From $N$, we compute $\dot n_{\gamma, \, \rm{IC}}$ and $n_\gamma$. We note that as obvious from equation 1 in \citetads{2021A&A...646A.115W}, $n_\gamma$ is the sum of the two contributions $n_{\gamma,\,\mathrm{i}}$, i.e. the photon injection rate attenuated by the loss rate, and $n_{\gamma,\,\mathrm{IC}}$, i.e. the IC photon production rate attenuated by the loss rate. 

As the flux of injected particles might be well collimated, the escaping photons are assumed to leave the interaction region also along a beam of small opening angle $\phi$. After leaving the AGN, we assume propagation without pair absorption due to the infrared \added{(IR)} torus and extragalactic background light which becomes relevant only at higher energies \citepads{2008Sci...320.1752M}. Then, the detected flux density $F_{\rm{casc}}(x_\gamma)$ \added{in the AGN frame }is determined from $n_\gamma$ by equation 3 of \citetads{2021A&A...646A.115W}.\added{
%In SSC or EC models, relativistic electrons with an isotropic distribution are considered in an isotropically magnetized blob, which itself moves relativistically along the jet. Therefore, the escaping spectrum has to be transformed from the comoving blob frame to the AGN frame, leading to Doppler-boosting. 
%In contrast, in our model, electrons and photons stream away from the central object along the jet. An isotropic, relativistically moving blob hasn't formed yet. Hence, $N$ and $n_\gamma$ are measured in the AGN frame. Therefore, we do not perform an additional transformation, in line with \citetads{2021A&A...646A.115W}.
By contrast to models in which relativistically moving blobs are assumed, a beam of pairs and photons propagating away from the central object along the jet is assumed here \citepads{2021A&A...646A.115W}. Hence, $N$ and $n_\gamma$ are measured in the AGN frame.}
To show that the gamma-ray emission of 3C\,279 originates from the BLR, it is sufficient to find the parameters of equations \ref{EquationHEPInjectionScenario1} - \ref{EquationDistributionLEPs}, as well as \replaced{$R$}{$l$} and $\phi$, such that $F_{\rm{casc}}$ meets the observed flux density.

\section{Results and conclusions} \label{SectionResults}

\begin{figure*}
\parbox{.41\linewidth}{
 \captionof{table}{\\Input Parameters, Used to Fit the SED, and Attained $\chi_{\rm{red}}^2$}
\begin{flushleft}\vspace{-0.5cm}
 \begin{tabular}{| E{0.09\columnwidth} | E{0.08\columnwidth} | E{0.08\columnwidth} | E{0.08\columnwidth} |} % centered columns
 \hline
 \multirow{2}{*}{Quantity}	& \multicolumn{3}{c|}{Used Value}											\\
     								\phn	& Case 1\textsubscript{a} & Case 2 & Case 3 \\ \hline
 $x_{\gamma, \, 1}$	                    & $2.0 \times 10^2$					& $2.0 \times 10^2$			&	\phn				\\ \arrayrulecolor{lightgray}\hline\arrayrulecolor{black}
 $x_{\gamma, \, 0}$	                    & $9.8 \times 10^{5}$			    & $9.8 \times 10^{5}$		&	\phn					\\ \arrayrulecolor{lightgray}\hline\arrayrulecolor{black}
 $\gamma_{{\rm{i}},\,1}$	                & $\gamma_{\rm{mean}}-3 \varsigma$  & \phn                         	& $98$	\\ \arrayrulecolor{lightgray}\hline\arrayrulecolor{black}
 $\gamma_{{\rm{i}},\,0}$                 & $\gamma_{\rm{mean}}+3 \varsigma$  & \phn                         	& $9.8 \times 10^{5}$	\\ \arrayrulecolor{lightgray}\hline\arrayrulecolor{black}
 $\phi \, \, \, [\degr]$						& $2.0$					& $2.0$					& $2.0$	\\ \arrayrulecolor{lightgray}\hline\arrayrulecolor{black}
 $l \, \, \, [\rm{m}]$							& $2.4 \times 10^{15}$	& $2.4 \times 10^{15}$	& $2.4 \times 10^{15}$	\\ \arrayrulecolor{lightgray}\hline\arrayrulecolor{black}
 $K_{\rm{lines}} \, \, \, [\rm{m^{-3}}]$			& $1.2 \times 10^9$ 	& $1.1 \times 10^9$		& $1.2 \times 10^9$	\\ \arrayrulecolor{lightgray}\hline\arrayrulecolor{black}
 $K_{\rm{G}} \, \, \, [\rm{s^{-1} m^{-3}}]$		& $5.8 \times 10^{-3}$	&	\phn					& $6.5 \times 10^{-3}$ \\ \arrayrulecolor{lightgray}\hline\arrayrulecolor{black}
 $K_{\rm{P}} \, \, \, [\rm{s^{-1} m^{-3}}]$		& $0.012$               & $0.014$               & $8.9 \times 10^3$	\\ \arrayrulecolor{lightgray}\hline\arrayrulecolor{black}
 $\gamma_{\rm{mean}}$                	& $3.9 \times 10^{5}$	& \phn						& $3.9 \times 10^{5}$	\\ \arrayrulecolor{lightgray}\hline\arrayrulecolor{black}
 $\varsigma / \gamma_{\rm{mean}}$		& $0.20$				& \phn  					& $0.20$							\\ \arrayrulecolor{lightgray}\hline\arrayrulecolor{black}
 $\alpha$								& $-2.1$				& $-2.2$				& $-2.3$			\\ \arrayrulecolor{lightgray}\hline\arrayrulecolor{black}
 $\beta$									&	\phn					& $0.068$				&	\phn							\\ \arrayrulecolor{lightgray}\hline\arrayrulecolor{black}
 $\chi_{\rm{red}}^2$                             & $0.90$                            & $0.62$                & $0.83$             \\
 \hline
 \end{tabular}
\end{flushleft}
 \label{TableParameters}\vspace{-0.5cm}
\tablecomments{For the meaning of the quantities, see the main text or \citetads{2021A&A...646A.115W}.}
}
\hfill
\parbox{.56\linewidth}{
\includegraphics[width=0.56\textwidth]{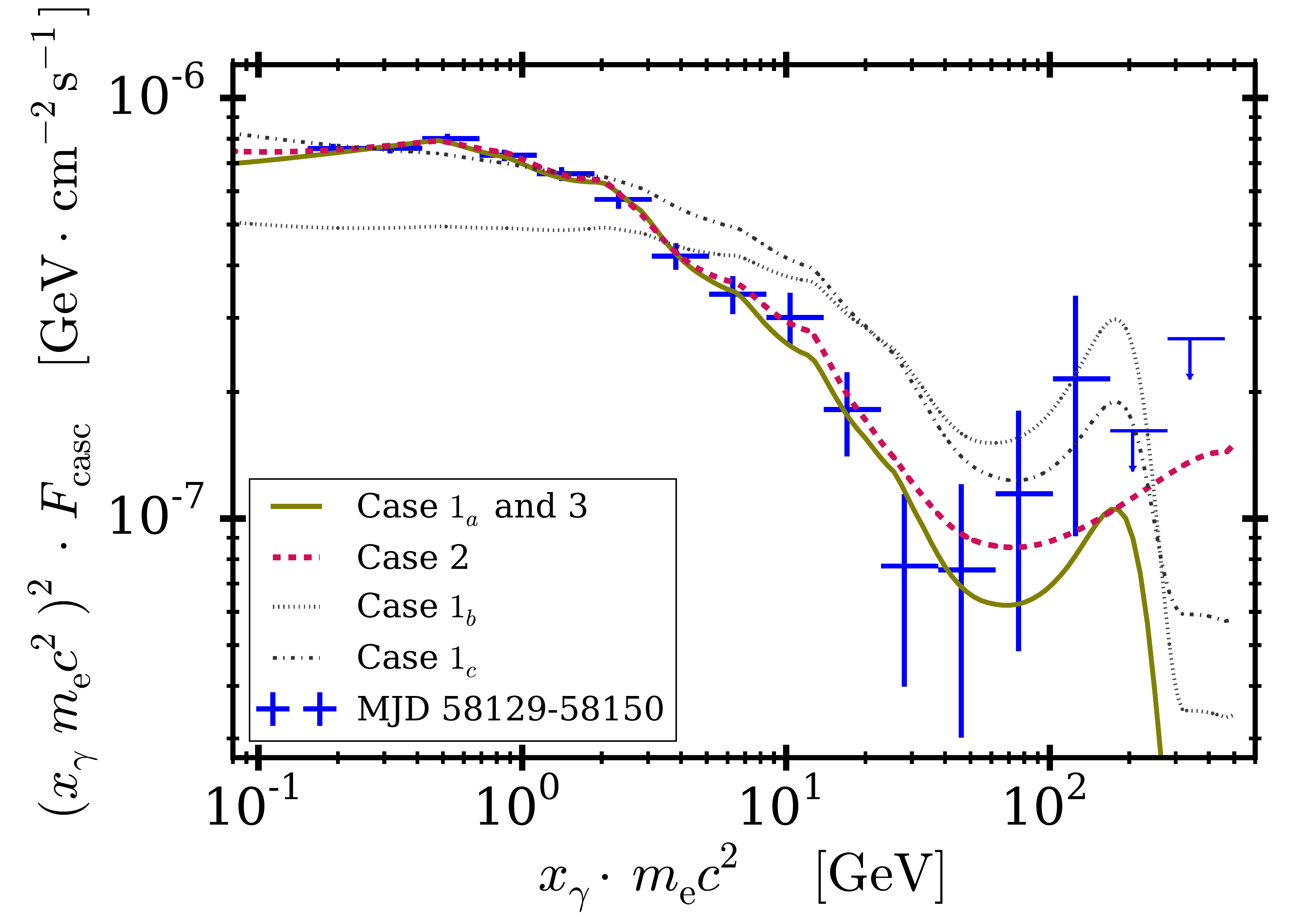}
\caption{Energy flux density vs. HE photon energy for the\deleted{ injection} cases considered as well as for the Fermi LAT observation. Cases \replaced{1a}{1\textsubscript{a}} and 3 are indistinguishable within drawing accuracy. The olive solid and the purple dashed curves depict cascades in the BLR field, while the gray dotted and dashed-dotted ones depict cascades on the diluted BLR field.\added{ The photon energy is measured in the AGN frame.}}
\label{FigureFluxDensity}
}
\end{figure*}

In \replaced{all three cases (namely case 1a, case 2 and case 3)}{the cases 1\textsubscript{a}, 2, and 3}, the observations can be fitted with our theoretical model by choosing appropriate input parameters yielding the acceptable reduced chi-squares $\chi_{\rm{red}}^2$ listed in table \ref{TableParameters}. Conventional least squares fits with log-parabolic or PL with exponential cutoff models yield $\chi_{\rm{red}}^2$s of 1.2 or 3.3, respectively. We note that the cutoff parameters $x_{\gamma, \, 1}$, $x_{\gamma, \, 0}$, $\gamma_{{\rm{i}},\,1}$, and $\gamma_{{\rm{i}},\,0}$ were not obtained by fitting, but were chosen to lie below and above the Fermi sensitivity range.

For the BLR spectrum we choose emission lines from the MUV, far-UV, extreme-UV (EUV), and soft X-ray regime as shown in table \ref{TableBLRLines}. We include lines that are prominent in observational BLR spectra and in synthetic spectra. Strong lines which are neighboring in $x$ by less than $\approx 10 \, \%$ are treated as one single line. We treat the numbers $K_{{\rm{line}},\,i}$ as parameters free within reasonable borders. We choose $K_{{\rm{line}},\,i}$ such that the tips in the pair-absorption rate (see figure \ref{FigureLossRate}) create troughs in the cascaded spectrum so that the observed SED is met.

 \begin{table*}
 \caption{Emission Lines, Used as the Soft Target Photons}
 \begin{tabular}
 {| r | c | r@{\hspace{0.4cm}} r@{}l | N{1.6cm} | N{1.6cm} | N{1.6cm} |} % centered columns (6 columns)
 \hline
 \multirow{2}{*}{$i\,\,$}	& \multirow{2}{*}{Line Designation} 	& \multicolumn{3}{c|}{Wavelength $\lambda_{0,\,i}$}	& \multicolumn{3}{c|}{Relative Flux Density Contribution $K_{{\rm{line}},\,i}$}		 \\
 								{}			&	{} & \multicolumn{3}{c|}{[$\rm{nm}$]} & Case 1\textsubscript{a} & Case 2 & Case 3 \\
 \hline
 % Scenario 1: Data from 188. run and "Broad emission lines from 2020-03-24 09-00.csv"
 % Scenario 2: Data from 189. run and "Broad emission lines from 2020-03-21 13-00.csv"
 % Scenario 3: Data from 193. run and "Broad emission lines from 2020-09-07 11-00.csv"
 1	& O VII \textsuperscript{a, b, c}                       & & 2.&20  & 5.90 & 4.80 & 5.20	\\ \arrayrulecolor{lightgray}\hline\arrayrulecolor{black}
 2	& C V \textsuperscript{a, b}                            & & 4.&05  & 3.55 & 4.25 & 3.45	\\ \arrayrulecolor{lightgray}\hline\arrayrulecolor{black}
 3	& Fe XVIII \textsuperscript{b, c}                       & & 9.&39  & 4.70 & 5.75 & 5.05	\\ \arrayrulecolor{lightgray}\hline\arrayrulecolor{black}
 4	& Fe XXIII \textsuperscript{b, c}                       & & 13.&3  & 3.95 & 2.90 & 3.80	\\ \arrayrulecolor{lightgray}\hline\arrayrulecolor{black}
 5	& He II Lyman continuum \textsuperscript{a}             & & 22.&8  & 0.75 & 0.45 & 0.70	\\ \arrayrulecolor{lightgray}\hline\arrayrulecolor{black}
 6	& He II Lyman-$\alpha$ \textsuperscript{a, b, c}        & & 30.&5  & 3.70 & 2.90 & 3.50	\\ \arrayrulecolor{lightgray}\hline\arrayrulecolor{black}
 7	& He I \textsuperscript{a, b, c}                        & & 58.&4  & 4.80 & 4.75 & 5.30 \\ \arrayrulecolor{lightgray}\hline\arrayrulecolor{black}
 8	& H I Lyman continuum\textsuperscript{e}                & & 93.&0  & 1.75 & 1.60 & 1.75	\\ \arrayrulecolor{lightgray}\hline\arrayrulecolor{black}
 9	& H I Lyman-$\alpha$ \textsuperscript{a, b, c, d, e}    & & 122\,& & 4.35 & 1.70 & 4.35	\\ \arrayrulecolor{lightgray}\hline\arrayrulecolor{black}
 10	& C IV \textsuperscript{a, b, c, d, e}                  & & 155\,& & 1.85 & 0.70 & 1.90 \\ \arrayrulecolor{lightgray}\hline\arrayrulecolor{black}
 11	& Mg II \textsuperscript{b, c, d, e}                    & & 280\,& & 0.55 & 0.20 & 0.60	\\
 \hline
 \end{tabular}
 \label{TableBLRLines}
 \tablecomments{The lines labeled by letters are prominent in the corresponding reference:
 a: \citetads{2017MNRAS.464..152A}, % Abolmasov
 b: atomic and molecular physics database by R.L. Kelly from Harvard-Smithsonian Center for Astrophysics, 
 c: Chianti atomic database for astrophysical spectroscopy by \citetads{1997A&AS..125..149D} and  \citetads{2012ApJ...744...99L}, 
 d: \citetads{2002ApJ...565..773T}, % Telfer
 e: \citetads{2005MNRAS.361..919P}, % Pian
 % f: ATOMDB Atomic data for astrophysicists version 3.0.9 \citepads[cf.][]{2010AAS...21547607F}, 
 }
 \end{table*}
%%%%%%%%%%%%%%%%%%%%%%%%%%%%% End of an old table fitting but not ideal for ApJL %%%%%%%%%%%%%%%%%%%%%%%%%%

{\catcode`\&=11 \gdef\1997AandAS..125..149D{\citetads{1997A&AS..125..149D}}} % This is necessary so that citations with an "&" can be used in tables.

As can be seen in table \ref{TableBLRLines}, the $K_{{\rm{line}},\,i}$ in the soft X-ray regime are of similar order of magnitude as the $K_{{\rm{line}},\,i}$ in the UV regime. This is the case because in our approach the numbers $K_{{\rm{line}},\,i}$ are not to be understood as a proxy for the equivalent widths of the lines. Typical BLR spectra consist of emission lines and of continuum emission reflected from the illuminating accretion disk and X-ray-emitting corona \citepads[see][]{2008MNRAS.386..945T}. This continuum emission is also included in the coefficients $K_{{\rm{line}},\,i}$. 
%Such high disk temperatures are accessible only in the very inner portion of the AGN, pointing to the hypothesis of the cascade taking place in the very interior.

We show the spectral pair-absorption rate of case \replaced{1a}{1\textsubscript{a}} as well as the corresponding optical depth in figure \ref{FigureLossRate}. The corresponding plots for cases 2 and 3 look qualitatively similar. The course of our optical depth with energy from $10 \, \rm{GeV}$ up to $\approx 90 \, \rm{GeV}$ is similar (slightly larger in comparison) to the one of \citetads{2006ApJ...653.1089L} with assuming the gamma-ray-emitting region at the inner edge of the BLR (dashed line in their figure 8). Between $\approx 30 \, \rm{GeV}$ and $\approx 90 \, \rm{GeV}$ our optical depth is similar to the one by the \citetads{2019A&A...627A.159H} with assuming a shell-like BLR and emission from slightly within the BLR (cyan curves in the left-hand panel of their figure 5) or with assuming a ring-like BLR and emission from deeply within the BLR (blue curves in the right-hand panel of their figure 5). In contrast to \citetads{2006ApJ...653.1089L} and the \citetads{2019A&A...627A.159H}, in our case inclusion of EUV and soft X-ray lines results in extension of the optical depth to energies below $1 \, \rm{GeV}$. Above $\approx 90 \, \rm{GeV}$, our optical depth decreases with energy due to the neglected NUV and optical lines. \citetads{2009MNRAS.392L..40T} and the \citetads{2019A&A...627A.159H} have stronger optical to \replaced{infrared}{IR} contribution in their target spectra and thus higher optical depths above $\approx 100 \, \rm{GeV}$. In comparison to the shell-like BLR model of \citetads{2012arXiv1209.2291T} with comparable disk luminosity, our optical depth is higher below $\approx 10 \, \rm{GeV}$, but above that it is of the same order of magnitude.

%%%%%%%%%%%%%% Figure for ApJ (Begin) %%%%%%%%%%%%%%
%\begin{figure}
%\resizebox{\hsize}{!}{\includegraphics{C:/Chris's/PhyMaLaChePyC/Physik/Astro - Modelling 3C279/n times xgamma^2 versus final photon energy from 2020-09-02 12-46.pdf}}
%%\resizebox{\hsize}{!}{\includegraphics{n times xgamma^2 versus final photon energy from 2020-09-02 12-46.pdf}}
%\caption{Product of the squared HE photon energy with the spectral number density of photons (solid) as well as with its two contributions $n_{\gamma,\,\mathrm{i}}$ (dashed) and $n_{\gamma,\,\mathrm{IC}}$ (dotted) vs. the product of the HE photon energy with the energy of the highest energetic line (bottom abscissa) and vs. the HE photon energy in GeV (top abscissa). All curves are for case \replaced{1a}{1\textsubscript{a}}. For the exact meaning of the plotted quantities, see section \ref{SectionModeling} or \citetads{2021A&A...646A.115W}.}
%\label{FigureHEPDistribution2}
%\end{figure}
%%%%%%%%%%%%%% Figure for ApJ (End) %%%%%%%%%%%%%%%%

We also show the photon injection rate $\dot n_{\gamma , \, \rm i}$, the production rate $\dot n_{\gamma, \, \rm{IC}}$ as well as their sum $\dot n_{\gamma}$ exemplarily for case \replaced{1a}{1\textsubscript{a}} in figure \ref{FigureHEPDistribution1}. In case \replaced{1a}{1\textsubscript{a}}, at energies below $\approx 2 \, \rm{GeV}$, the emerging radiation is composed about equally from $\dot n_{\gamma,\,\mathrm{i}}$ and from $\dot n_{\gamma,\,\mathrm{IC}}$. With increasing energy, the contribution of the injected photons decreases, and that one of the IC up-scattered photons increases. In case 2, IC-produced photons dominate the photon population between $\approx 0.4 \, \rm{GeV}$ and $\approx 100 \, \rm{GeV}$, while as a consequence of the curvature of the log-parabola the injected photons dominate elsewhere. In case 3, there is no photon injection and consequently the entire photon population is due to cascaded radiation.

The photon number density $n_{\gamma}$ and its two contributions $n_{\gamma , \, \rm i}$ (i.e. injection rate divided by loss rate) and $n_{\gamma, \, \rm{IC}}$ (i.e. production rate divided by loss rate) are shown for case \replaced{1a}{1\textsubscript{a}} in figure \ref{FigureHEPDistribution2}. There, the absorption features by the emission lines are seen as the dips above $\approx 0.5 \, \rm{GeV}$. The major dip above $\approx 20 \, \rm{GeV}$ is caused in all \deleted{three }cases by pair absorption on He II, He I and H I emission line photons.

The resulting energy flux densities in comparison to the observational data are shown in figure \ref{FigureFluxDensity}. The bump in the SEDs above $\approx 100 \, \rm{GeV}$ is in cases \replaced{1a}{1\textsubscript{a}} and 3 the result of strong electron injection around $\approx 200 \, \rm{GeV}$, and in case 2 due to the log-parabolic photon injection. The fact that the modeled flux densities can describe the observational ones in \replaced{all three cases}{the cases 1\textsubscript{a}, 2 and 3} means that HE gamma-ray emission from the edge of the BLR is a robust finding and especially independent of the precise composition of the injected species.
% A qualitatively similar trough would appear in the SED via absorption of gamma rays in a stellar radiation field \citepads{2021MNRAS.tmp..561B}. This would however necessitate an extreme stellar surface temperature to facilitate pair absorption on EUV and X-ray photons}
Interactions of gamma rays with optical emission line photons of stellar radiation fields can produce absorption troughs lacking, however, the features due to EUV or X-ray lines present in the BLR radiation field \citepads{2021MNRAS.tmp..561B}.

To estimate the total luminosity of the BLR for case \replaced{1a}{1\textsubscript{a}}, we determine the total energy density of the soft photons through $u_{\rm{BLR,\,tot}} = K_{\rm{lines}} \cdot \sum _{i=1}^{11} K_{{\rm{line}},\,i} \cdot m_{\rm e} c^2$. From this we get the luminosity \replaced{$L_{\rm{BLR,\,tot}} = u_{\rm{BLR,\,tot}} \cdot c \cdot 4 \pi R^2 = 7.4 \times 10^{37} \, \rm{W}$}{$L_{\rm{BLR,\,tot}} = u_{\rm{BLR,\,tot}} \cdot c \cdot 4 \pi l^2 = 7.4 \times 10^{37} \, \rm{W}$}. Analogously, we determine the UV luminosity \replaced{$L_{\rm{BLR,\,UV}} = K_{\rm{lines}} \cdot \sum _{i=4}^{11} K_{{\rm{line}},\,i} \cdot m_{\rm e} c^2 \cdot c \cdot 4 \pi R^2 = 4.5 \times 10^{37} \, \rm{W}$}{$L_{\rm{BLR,\,UV}} = K_{\rm{lines}} \cdot \sum _{i=4}^{11} K_{{\rm{line}},\,i} \cdot m_{\rm e} c^2 \cdot c \cdot 4 \pi l^2 = 4.5 \times 10^{37} \, \rm{W}$}. From this we get an estimate of the BLR radius %$R_{\rm BLR}$ 
with help of the empirical relation by \citetads[][equation 3 therein]{2007ApJ...659..997K}, connecting the time-lag-based C IV radius with the UV luminosity. We obtain $R_{\rm C \, IV} = 4.5 \times 10^{14} \, \rm m$. Considering the scatter of the normalizations in empirical radius luminosity relations (\citeads{2008MNRAS.387.1669G}; \citeads{2015ApJ...801....8K}),
%and taking into account that the outer border $R_{\rm BLR}$ of the BLR is bigger than $R_{\rm C \, IV}$ by a factor of a few \citepads[e.g.][]{2020MNRAS.494.1611N}
the radii $R_{\rm C \, IV}$ and \replaced{$R$}{$l$} can be considered similar. For example, using the canonical $10 \, \%$ reprocessing fraction \citepads{2017MNRAS.469..255G}, the disk luminosity is $L_{\rm{disk,\,tot}} = 10 \, L_{\rm{BLR,\,tot}}$ and the approximate BLR radius is $2.7 \times 10^{15} \, \rm{m}$ \citepads[][equation 4 therein]{2008MNRAS.387.1669G}.

%%%%%%%%%%%%%% Figure for ApJ (Begin) %%%%%%%%%%%%%%
%\begin{figure}
%\resizebox{\hsize}{!}{\includegraphics{C:/Chris's/PhyMaLaChePyC/Physik/Astro - Modelling 3C279/Energy-flux-density versus photon energy from 2021-03-17 19-30-3.pdf}}
%%\resizebox{\hsize}{!}{\includegraphics{Energy-flux-density versus photon energy from 2021-03-17 19-30-3.pdf}}
%\caption{Energy flux density vs. HE photon energy for the\deleted{ injection} cases considered as well as for the Fermi LAT observation. Cases \replaced{1a}{1\textsubscript{a}} and 3 are indistinguishable within drawing accuracy. The olive solid and the purple dashed curves depict cascades in the BLR field, while the gray dotted and dashed-dotted ones depict cascades on the diluted BLR field.\added{ The photon energy is measured in the AGN frame.}}
%\label{FigureFluxDensity}
%\end{figure}
%%%%%%%%%%%%%% Figure for ApJ (End) %%%%%%%%%%%%%%%%

Now, we argue that emission from outside of the BLR is hardly realizable within our model and the Fermi SED. Modeling a cascade on the diluted BLR photon field outside of the BLR would correspond to a reduced soft photon density, reflected in a reduced $K_{\rm{lines}}$\added{, see figure \ref{FigureSketch}}. Exemplarily, for the \replaced{case 1 injection}{injection equation \ref{EquationHEPInjectionScenario1} and \ref{EquationElectronInjectionScenario1} (case 1 injection)} we choose $K_{\rm{lines}}$ being $0.25$ of the value reported in table \ref{TableParameters}, corresponding to dilution of radiation at a doubled distance\added{, and denote this case 1\textsubscript{b}}. The interaction radius is kept the same as in table \ref{TableParameters}. This means that the injected particles are injected at the outer edge of the BLR and interact with the diluted photon field. If they were injected inside, they would be reprocessed inside, as happening in cases 1\textsubscript{a}, 2, and 3. With this reduced $K_{\rm{lines}}$, we obtain the SED depicted by case 1\textsubscript{b} in figure \ref{FigureFluxDensity} with $\chi_{\rm{red}}^2 = 106$. We run a series of simulations \added{by changing the parameters of equation \ref{EquationHEPInjectionScenario1} and \ref{EquationElectronInjectionScenario1} }and try to find a new set of parameters to fit the Fermi data again with this reduced $K_{\rm{lines}}$.
%%%%%%%%%%%%%% The following lines were substituted due to being verbose. %%%%%%%%%%%%%%
% We cannot find a satisfactory fit, what can be understood as follows. The reduction of $K_{\rm{lines}}$ causes less pair absorption and a stronger influence of escape. This leads to an increased flux density above $\approx 5 \, \rm GeV$ and to a decreased flux density below $\approx 3 \, \rm GeV$ (cf. gray dotted line in figure \ref{FigureFluxDensity}). Then, to meet the data points below $3 \, \rm GeV$, the normalization $K_{\rm P}$ has to be increased. This (as well as a gentle decrease of $K_{\rm G}$ to not overshoot the points around 100 GeV) is done in the case 1\textsubscript{c} SED in figure \ref{FigureFluxDensity}, yielding $\chi_{\rm{red}}^2 = 12.8$. However, this leads to an increasing contribution of $\dot n_{\gamma , \, \rm i}$ to the flux density, resulting in too soft a slope below $1 \, \rm GeV$. We could mitigate this by increasing $\alpha$, which however would increase the flux density far over the observational points in the range around $10 \, \rm GeV$. To meet these points again, one had two possibilities. First, one could almost entirely remove the electron injection, and thus reduce $\dot n_{\gamma , \, \rm IC}$. Then, however, we would not meet the points around 100 GeV. Second, one could increase pair absorption around $10 \, \rm GeV$, by increasing the corresponding $K_{{\rm{line}},\,i}$. This would however raise the BLR luminosity over reasonable values.
%%%%%%%%%%%%% The preceding lines were substituted by the following lines: %%%%%%%%%%%%%
Without raising the BLR luminosity over reasonable values, we do not achieve a model fit significantly better than the case 1\textsubscript{c} in figure \ref{FigureFluxDensity} with $\chi_{\rm{red}}^2 = 12.8$. The slope below $1 \, \rm GeV$ cannot be reconciled with the trough above $\approx 3 \, \rm GeV$.

{\catcode`\&=11 \gdef\2021AandA...646A.115W{\citetads{2021A&A...646A.115W}}} % This is necessary so that citations with an "&" can be used in tables.
%%%%%%%%%%%%%% Table and Figure for arXiv (End) %%%%%%%%%%%%%%%%

\section{Summary}\label{SectionSummary}

%\added{SSC and EC models involving relativistically moving blobs can explain SEDs consisting of an IR bump from synchrotron radiation and a HE bump due to IC-scattering. Additional cascaded emission from in or near the BLR can account for intermediate-time- variability as well as for substructure on top of the HE bump \citepads{2021A&A...646A.115W}. }
\replaced{We have considered the possibility that emission from an IC pair cascade is responsible for the HE gamma rays detected by the Fermi LAT from 3C 279.}{It is generally accepted that lepto-hadronic emission models can describe the nonthermal SEDs of blazars such as 3C\,279 based on the assumption of particle acceleration at shock waves traveling down the jets. The high amplitudes and short timescales of the gamma-ray flux variations, however, indicate the injection of major dissipative events showing temporal and spectral features that call for explanations beyond the shock acceleration scenario (\citeads{2020NatCo..11.4176S}; \citeads{2021A&A...646A.115W}).

Here, we have shown that the gamma-ray spectrum of 3C\,279 measured with high precision in an active state is in agreement with an IC pair cascade spectrum initiated by a beam of pairs or gamma rays, including absorption troughs from the interactions of the gamma rays and pairs with emission line photons from the edge of the BLR.}

\added{Below MeV energies, the reprocessing of cascaded emission is limited due to the dominating impact of escape. To account for the SED in a wider energy range, it is necessary to involve additional radiation processes (such as synchrotron emission in transverse magnetic fields) or additional emission sites (e.g. shock waves traveling down the jet).}

We \replaced{suggested}{investigated} three possible cases that differ mainly in the functional shape of the injected species.
%We inject PL photons plus Gaussian electrons in scenario 1, logparabolic photons in scenario 2, and PL plus Gaussian electrons in scenario 3.
In all three cases we \replaced{used BLR radiation as target photons. This is equivalent with the cascade happening on the BLR field.}{considered the radiation field of the BLR region as a target.} The \replaced{escaping}{emerging} spectra of the cascade were computed numerically and are superpositions of the injected spectra attenuated by pair absorption and \replaced{by}{of} IC up-scattered emission. We achieved fits to the data in all three cases \added{1\textsubscript{a}, 2, and 3, }and found the BLR UV-luminosity-based radius being of the same order of magnitude as the radial size of the cascade region\added{, which itself is in agreement with estimates on the radius of the $\gamma$-sphere of 3C\,279 at 1\,GeV \citepads{1995ApJ...441...79B}}.

%%%%%%%%%%% The following is from an earlier version, about from autumn 2020. %%%%%%%%%%%%
%A cascade driven by radiation from outside of the BLR can be modeled with usage of a lowered soft photon field and the same size of the interaction region. Within this prerequisite we did not succeed in finding reasonable fits to the data. We interpret this as evidence for emission of HE gamma rays from within the BLR. Of course, as multiple emitting regions can exist simultaneously and can propagate along the jet, evidence for a certain emission location does not exclude emission at one or more different locations, especially during different times or for other objects (\citeads{2011MNRAS.417L..11S}; \citeads{2014ApJ...794....8S}; \citeads{2015PASJ...67...79L}; \citeads{2019A&A...630A..56P}; \citeads{2020NatCo..11.4176S}).
%%%%%%%%%%% The preceding is from an earlier version, about from autumn 2020. %%%%%%%%%%%%

A cascade taking place outside of the BLR was modeled \added{(with cases 1\textsubscript{b} and 1\textsubscript{c})} by choosing a lower density of the soft photon field and the same size of the interaction region (escape timescale). We found, however, that this precludes us from achieving satisfactory fits to the data, strengthening our interpretation of the gamma-ray SED observed with Fermi LAT as evidence for emission from the edge of the BLR.

%The beam particles in the spine of the jet could originate from the activity  of the central engine, propagating to the edge of the BLR protected by the sheath, or from particle acceleration associated with structural changes of the jet leaving the BLR
If a sheath surrounding the jet blocks the BLR photons, the cascade could be initiated by particles accelerated much deeper within the BLR. Moreover, multiple emitting regions could exist simultaneously and could propagate along the jet (\citeads{2011MNRAS.417L..11S}; \citeads{2014ApJ...794....8S};
\citeads{2014A&A...567A..41A}; \citeads{2015PASJ...67...79L}; \citeads{2016ApJ...830...94F}; \citeads{2018ApJ...858...80R}; \citeads{2019A&A...630A..56P}; \citeads{2020NatCo..11.4176S}; \citeads{2021MNRAS.500.5297A}). 
It may therefore be difficult to generalize our conclusions, unless there is a causal connection between the gamma-ray emitting region and the BLR.

%\added{\\Below $\approx 0.1 \, \rm{GeV}$, the emission of the cascade considered in our work decreases with decreasing energy. This is because of the limitation of radiation reprocession in this energy regime due to the dominating impact of escape. Therefore, our model cannot account for emission below several $10 \, \rm{MeV}$. Still, SSC / EC emission is necessary to account for the broadband two-bump blazar SED structure. Hence our model is no substitute of common SSC or EC models, but an addendum to account for the fine structure on top of the beamed jet emission. Karl:  Moreover, to account for the SED below 100\,MeV and above 100\,GeV, it is necessary to assume additional radiation processes (such as synchrotron emission, if the pair beam encounters magnetic fields perpendicular to the direction of its propagation) or additional emission sites (e.g. shock waves traveling down the jet).}

\acknowledgments
We are grateful to the developers of NumPy \citep{10.5555/2886196} and matplotlib \citepads{2007CSE.....9...90H} as well as to sensible comments from the referee. Chianti is a collaborative project involving the George Mason University, the University of Michigan, the University of Cambridge and the NASA Goddard Space Flight Center (GSFC). We thank the Fermi LAT Collaboration and the GSFC for providing LAT data and software to analyze the LAT data. C.W. acknowledges support by the German Bundesministerium f\"{u}r Arbeit und Soziales and by "Promotion inklusive" of the Universit\"{a}t zu K\"{o}ln.

\facility{Fermi (LAT).}

\bibliographystyle{aasjournal} % style aasjournal.bst
\bibliography{2021-07-30-ApJ-paper}{} % your references Yourfile.bib

\begin{thebibliography}{}
\expandafter\ifx\csname natexlab\endcsname\relax\def\natexlab#1{#1}\fi
\providecommand{\url}[1]{\href{#1}{#1}}
\providecommand{\dodoi}[1]{doi:~\href{http://doi.org/#1}{\nolinkurl{#1}}}
\providecommand{\doeprint}[1]{\href{http://ascl.net/#1}{\nolinkurl{http://ascl.net/#1}}}
\providecommand{\doarXiv}[1]{\href{https://arxiv.org/abs/#1}{\nolinkurl{https://arxiv.org/abs/#1}}}

\bibitem[{{Abolmasov} \& {Poutanen}(2017)}]{2017MNRAS.464..152A}
{Abolmasov}, P., \& {Poutanen}, J. 2017, \mnras, 464, 152,
  \dodoi{10.1093/mnras/stw2326}

\bibitem[{{Acharyya} {et~al.}(2021){Acharyya}, {Chadwick}, \&
  {Brown}}]{2021MNRAS.500.5297A}
{Acharyya}, A., {Chadwick}, P.~M., \& {Brown}, A.~M. 2021, \mnras, 500, 5297,
  \dodoi{10.1093/mnras/staa3483}

\bibitem[{{Ackermann} {et~al.}(2016){Ackermann}, {Anantua}, {Asano}, {Baldini},
  {Barbiellini}, {Bastieri}, {Becerra Gonzalez}, {Bellazzini}, {Bissaldi},
  {Blandford}, {Bloom}, {Bonino}, {Bottacini}, {Bruel}, {Buehler}, {Caliandro},
  {Cameron}, {Caragiulo}, {Caraveo}, {Cavazzuti}, {Cecchi}, {Cheung}, {Chiang},
  {Chiaro}, {Ciprini}, {Cohen-Tanugi}, {Costanza}, {Cutini}, {D'Ammando}, {de
  Palma}, {Desiante}, {Digel}, {Di Lalla}, {Di Mauro}, {Di Venere}, {Drell},
  {Favuzzi}, {Fegan}, {Ferrara}, {Fukazawa}, {Funk}, {Fusco}, {Gargano},
  {Gasparrini}, {Giglietto}, {Giordano}, {Giroletti}, {Grenier}, {Guillemot},
  {Guiriec}, {Hayashida}, {Hays}, {Horan}, {J{\'o}hannesson}, {Kensei},
  {Kocevski}, {Kuss}, {La Mura}, {Larsson}, {Latronico}, {Li}, {Longo},
  {Loparco}, {Lott}, {Lovellette}, {Lubrano}, {Madejski}, {Magill}, {Maldera},
  {Manfreda}, {Mayer}, {Mazziotta}, {Michelson}, {Mirabal}, {Mizuno},
  {Monzani}, {Morselli}, {Moskalenko}, {Nalewajko}, {Negro}, {Nuss}, {Ohsugi},
  {Orlando}, {Paneque}, {Perkins}, {Pesce-Rollins}, {Piron}, {Pivato},
  {Porter}, {Principe}, {Rando}, {Razzano}, {Razzaque}, {Reimer}, {Scargle},
  {Sgr{\`o}}, {Sikora}, {Simone}, {Siskind}, {Spada}, {Spinelli}, {Stawarz},
  {Thayer}, {Thompson}, {Torres}, {Troja}, {Uchiyama}, {Yuan}, \&
  {Zimmer}}]{2016ApJ...824L..20A}
{Ackermann}, M., {Anantua}, R., {Asano}, K., {et~al.} 2016, \apjl, 824, L20,
  \dodoi{10.3847/2041-8205/824/2/L20}

\bibitem[{{Aleksi{\'c}} {et~al.}(2014){Aleksi{\'c}}, {Ansoldi}, {Antonelli},
  {Antoranz}, {Babic}, {Bangale}, {Barres de Almeida}, {Barrio}, {Becerra
  Gonz{\'a}lez}, {Bednarek}, {Berger}, {Bernardini}, {Biland}, {Blanch},
  {Bock}, {Bonnefoy}, {Bonnoli}, {Borracci}, {Bretz}, {Carmona}, {Carosi},
  {Carreto Fidalgo}, {Colin}, {Colombo}, {Contreras}, {Cortina}, {Covino}, {Da
  Vela}, {Dazzi}, {De Angelis}, {De Caneva}, {De Lotto}, {Delgado Mendez},
  {Doert}, {Dom{\'\i}nguez}, {Dominis Prester}, {Dorner}, {Doro}, {Einecke},
  {Eisenacher}, {Elsaesser}, {Farina}, {Ferenc}, {Fonseca}, {Font}, {Frantzen},
  {Fruck}, {Garc{\'\i}a L{\'o}pez}, {Garczarczyk}, {Garrido Terrats}, {Gaug},
  {Giavitto}, {Godinovi{\'c}}, {Gonz{\'a}lez Mu{\~n}oz}, {Gozzini}, {Hadasch},
  {Herrero}, {Hildebrand}, {Hose}, {Hrupec}, {Idec}, {Kadenius}, {Kellermann},
  {Knoetig}, {Kodani}, {Konno}, {Krause}, {Kubo}, {Kushida}, {La Barbera},
  {Lelas}, {Lewandowska}, {Lindfors}, {Lombardi}, {L{\'o}pez},
  {L{\'o}pez-Coto}, {L{\'o}pez-Oramas}, {Lorenz}, {Lozano}, {Makariev},
  {Mallot}, {Maneva}, {Mankuzhiyil}, {Mannheim}, {Maraschi}, {Marcote},
  {Mariotti}, {Mart{\'\i}nez}, {Mazin}, {Menzel}, {Meucci}, {Miranda},
  {Mirzoyan}, {Moralejo}, {Munar-Adrover}, {Nakajima}, {Niedzwiecki},
  {Nilsson}, {Nishijima}, {Nowak}, {Orito}, {Overkemping}, {Paiano},
  {Palatiello}, {Paneque}, {Paoletti}, {Paredes}, {Paredes-Fortuny}, {Partini},
  {Persic}, {Prada}, {Prada Moroni}, {Prandini}, {Preziuso}, {Puljak},
  {Reinthal}, {Rhode}, {Rib{\'o}}, {Rico}, {Rodriguez Garcia}, {R{\"u}gamer},
  {Saggion}, {Saito}, {Saito}, {Salvati}, {Satalecka}, {Scalzotto}, {Scapin},
  {Schultz}, {Schweizer}, {Shore}, {Sillanp{\"a}{\"a}}, {Sitarek}, {Snidaric},
  {Sobczynska}, {Spanier}, {Stamatescu}, {Stamerra}, {Steinbring}, {Storz},
  {Sun}, {Suri{\'c}}, {Takalo}, {Takami}, {Tavecchio}, {Temnikov},
  {Terzi{\'c}}, {Tescaro}, {Teshima}, {Thaele}, {Tibolla}, {Torres}, {Toyama},
  {Treves}, {Vogler}, {Wagner}, {Zandanel}, \& {Zanin}}]{2014A&A...567A..41A}
{Aleksi{\'c}}, J., {Ansoldi}, S., {Antonelli}, L.~A., {et~al.} 2014, \aap, 567,
  A41, \dodoi{10.1051/0004-6361/201323036}

\bibitem[{{Appl} {et~al.}(2000){Appl}, {Lery}, \& {Baty}}]{2000A&A...355..818A}
{Appl}, S., {Lery}, T., \& {Baty}, H. 2000, \aap, 355, 818.
\newblock \url{https://ui.adsabs.harvard.edu/abs/2000A&A...355..818A}

\bibitem[{{Asano} \& {Hayashida}(2015)}]{2015ApJ...808L..18A}
{Asano}, K., \& {Hayashida}, M. 2015, \apjl, 808, L18,
  \dodoi{10.1088/2041-8205/808/1/L18}

\bibitem[{{Atwood} {et~al.}(2009){Atwood}, {Abdo}, {Ackermann}, {Althouse},
  {Anderson}, {Axelsson}, {Baldini}, {Ballet}, {Band}, {Barbiellini},
  {Bartelt}, {Bastieri}, {Baughman}, {Bechtol}, {B{\'e}d{\'e}r{\`e}de},
  {Bellardi}, {Bellazzini}, {Berenji}, {Bignami}, {Bisello}, {Bissaldi},
  {Blandford}, {Bloom}, {Bogart}, {Bonamente}, {Bonnell}, {Borgland},
  {Bouvier}, {Bregeon}, {Brez}, {Brigida}, {Bruel}, {Burnett}, {Busetto},
  {Caliandro}, {Cameron}, {Caraveo}, {Carius}, {Carlson}, {Casandjian},
  {Cavazzuti}, {Ceccanti}, {Cecchi}, {Charles}, {Chekhtman}, {Cheung},
  {Chiang}, {Chipaux}, {Cillis}, {Ciprini}, {Claus}, {Cohen-Tanugi},
  {Condamoor}, {Conrad}, {Corbet}, {Corucci}, {Costamante}, {Cutini}, {Davis},
  {Decotigny}, {DeKlotz}, {Dermer}, {de Angelis}, {Digel}, {do Couto e Silva},
  {Drell}, {Dubois}, {Dumora}, {Edmonds}, {Fabiani}, {Farnier}, {Favuzzi},
  {Flath}, {Fleury}, {Focke}, {Funk}, {Fusco}, {Gargano}, {Gasparrini},
  {Gehrels}, {Gentit}, {Germani}, {Giebels}, {Giglietto}, {Giommi}, {Giordano},
  {Glanzman}, {Godfrey}, {Grenier}, {Grondin}, {Grove}, {Guillemot}, {Guiriec},
  {Haller}, {Harding}, {Hart}, {Hays}, {Healey}, {Hirayama}, {Hjalmarsdotter},
  {Horn}, {Hughes}, {J{\'o}hannesson}, {Johansson}, {Johnson}, {Johnson},
  {Johnson}, {Johnson}, {Kamae}, {Katagiri}, {Kataoka}, {Kavelaars}, {Kawai},
  {Kelly}, {Kerr}, {Klamra}, {Kn{\"o}dlseder}, {Kocian}, {Komin}, {Kuehn},
  {Kuss}, {Landriu}, {Latronico}, {Lee}, {Lee}, {Lemoine-Goumard}, {Lionetto},
  {Longo}, {Loparco}, {Lott}, {Lovellette}, {Lubrano}, {Madejski}, {Makeev},
  {Marangelli}, {Massai}, {Mazziotta}, {McEnery}, {Menon}, {Meurer},
  {Michelson}, {Minuti}, {Mirizzi}, {Mitthumsiri}, {Mizuno}, {Moiseev},
  {Monte}, {Monzani}, {Moretti}, {Morselli}, {Moskalenko}, {Murgia},
  {Nakamori}, {Nishino}, {Nolan}, {Norris}, {Nuss}, {Ohno}, {Ohsugi}, {Omodei},
  {Orlando}, {Ormes}, {Paccagnella}, {Paneque}, {Panetta}, {Parent}, {Pearce},
  {Pepe}, {Perazzo}, {Pesce-Rollins}, {Picozza}, {Pieri}, {Pinchera}, {Piron},
  {Porter}, {Poupard}, {Rain{\`o}}, {Rando}, {Rapposelli}, {Razzano}, {Reimer},
  {Reimer}, {Reposeur}, {Reyes}, {Ritz}, {Rochester}, {Rodriguez}, {Romani},
  {Roth}, {Russell}, {Ryde}, {Sabatini}, {Sadrozinski}, {Sanchez}, {Sander},
  {Sapozhnikov}, {Parkinson}, {Scargle}, {Schalk}, {Scolieri}, {Sgr{\`o}},
  {Share}, {Shaw}, {Shimokawabe}, {Shrader}, {Sierpowska-Bartosik}, {Siskind},
  {Smith}, {Smith}, {Spandre}, {Spinelli}, {Starck}, {Stephens}, {Strickman},
  {Strong}, {Suson}, {Tajima}, {Takahashi}, {Takahashi}, {Tanaka}, {Tenze},
  {Tether}, {Thayer}, {Thayer}, {Thompson}, {Tibaldo}, {Tibolla}, {Torres},
  {Tosti}, {Tramacere}, {Turri}, {Usher}, {Vilchez}, {Vitale}, {Wang},
  {Watters}, {Winer}, {Wood}, {Ylinen}, \& {Ziegler}}]{2009ApJ...697.1071A}
{Atwood}, W.~B., {Abdo}, A.~A., {Ackermann}, M., {et~al.} 2009, \apj, 697,
  1071, \dodoi{10.1088/0004-637X/697/2/1071}

\bibitem[{{Baring} {et~al.}(2017){Baring}, {B{\"o}ttcher}, \&
  {Summerlin}}]{2017MNRAS.464.4875B}
{Baring}, M.~G., {B{\"o}ttcher}, M., \& {Summerlin}, E.~J. 2017, \mnras, 464,
  4875, \dodoi{10.1093/mnras/stw2344}

\bibitem[{{Bednarek} \& {Sitarek}(2021)}]{2021MNRAS.tmp..561B}
{Bednarek}, W., \& {Sitarek}, J. 2021, \mnras, 503, 2423,
  \dodoi{10.1093/mnras/stab554}

\bibitem[{{Blandford} {et~al.}(2019){Blandford}, {Meier}, \&
  {Readhead}}]{2019ARA&A..57..467B}
{Blandford}, R., {Meier}, D., \& {Readhead}, A. 2019, \araa, 57, 467,
  \dodoi{10.1146/annurev-astro-081817-051948}

\bibitem[{{Blandford} {et~al.}(2017){Blandford}, {Yuan}, {Hoshino}, \&
  {Sironi}}]{2017SSRv..207..291B}
{Blandford}, R., {Yuan}, Y., {Hoshino}, M., \& {Sironi}, L. 2017, \ssr, 207,
  291, \dodoi{10.1007/s11214-017-0376-2}

\bibitem[{{Blandford} \& {Levinson}(1995)}]{1995ApJ...441...79B}
{Blandford}, R.~D., \& {Levinson}, A. 1995, \apj, 441, 79,
  \dodoi{10.1086/175338}

\bibitem[{{B{\"o}ttcher} \& {Els}(2016)}]{2016ApJ...821..102B}
{B{\"o}ttcher}, M., \& {Els}, P. 2016, \apj, 821, 102,
  \dodoi{10.3847/0004-637X/821/2/102}

\bibitem[{{Britto} {et~al.}(2015){Britto}, {Razzaque}, \&
  {Lott}}]{2015arXiv150207624B}
{Britto}, R. J.~G., {Razzaque}, S., \& {Lott}, B. 2015, arXiv e-prints,
  arXiv:1502.07624.
\newblock \doarXiv{1502.07624}

\bibitem[{{Cash}(1979)}]{1979ApJ...228..939C}
{Cash}, W. 1979, \apj, 228, 939, \dodoi{10.1086/156922}

\bibitem[{{Dere} {et~al.}(1997){Dere}, {Landi}, {Mason}, {Monsignori Fossi}, \&
  {Young}}]{1997A&AS..125..149D}
{Dere}, K.~P., {Landi}, E., {Mason}, H.~E., {Monsignori Fossi}, B.~C., \&
  {Young}, P.~R. 1997, \aaps, 125, 149, \dodoi{10.1051/aas:1997368}

\bibitem[{{Dermer} {et~al.}(2014){Dermer}, {Cerruti}, {Lott}, {Boisson}, \&
  {Zech}}]{2014ApJ...782...82D}
{Dermer}, C.~D., {Cerruti}, M., {Lott}, B., {Boisson}, C., \& {Zech}, A. 2014,
  \apj, 782, 82, \dodoi{10.1088/0004-637X/782/2/82}

\bibitem[{{Donea} \& {Protheroe}(2003)}]{2003APh....18..377D}
{Donea}, A.-C., \& {Protheroe}, R.~J. 2003, Astroparticle Physics, 18, 377,
  \dodoi{10.1016/S0927-6505(02)00155-X}

\bibitem[{{Ebisuzaki} \& {Tajima}(2014)}]{2014APh....56....9E}
{Ebisuzaki}, T., \& {Tajima}, T. 2014, Astroparticle Physics, 56, 9,
  \dodoi{10.1016/j.astropartphys.2014.02.004}

\bibitem[{{Finke}(2016)}]{2016ApJ...830...94F}
{Finke}, J.~D. 2016, \apj, 830, 94, \dodoi{10.3847/0004-637X/830/2/94}

\bibitem[{{Ghisellini} {et~al.}(2017){Ghisellini}, {Righi}, {Costamante}, \&
  {Tavecchio}}]{2017MNRAS.469..255G}
{Ghisellini}, G., {Righi}, C., {Costamante}, L., \& {Tavecchio}, F. 2017,
  \mnras, 469, 255, \dodoi{10.1093/mnras/stx806}

\bibitem[{{Ghisellini} \& {Tavecchio}(2008)}]{2008MNRAS.387.1669G}
{Ghisellini}, G., \& {Tavecchio}, F. 2008, \mnras, 387, 1669,
  \dodoi{10.1111/j.1365-2966.2008.13360.x}

\bibitem[{{Giannios} \& {Spruit}(2006)}]{2006A&A...450..887G}
{Giannios}, D., \& {Spruit}, H.~C. 2006, \aap, 450, 887,
  \dodoi{10.1051/0004-6361:20054107}

\bibitem[{{Hartman} {et~al.}(1992){Hartman}, {Bertsch}, {Fichtel}, {Hunter},
  {Kanbach}, {Kniffen}, {Kwok}, {Lin}, {Mattox}, {Mayer-Hasselwand er},
  {Michelson}, {von Montigny}, {Nel}, {Nolan}, {Pinkau}, {Rothermel},
  {Schneid}, {Sommer}, {Sreekumar}, \& {Thompson}}]{1992ApJ...385L...1H}
{Hartman}, R.~C., {Bertsch}, D.~L., {Fichtel}, C.~E., {et~al.} 1992, \apjl,
  385, L1, \dodoi{10.1086/186263}

\bibitem[{{Hayashida} {et~al.}(2015){Hayashida}, {Nalewajko}, {Madejski},
  {Sikora}, {Itoh}, {Ajello}, {Blandford}, {Buson}, {Chiang}, {Fukazawa},
  {Furniss}, {Urry}, {Hasan}, {Harrison}, {Alexand er}, {Balokovi{\'c}},
  {Barret}, {Boggs}, {Christensen}, {Craig}, {Forster}, {Giommi},
  {Grefenstette}, {Hailey}, {Hornstrup}, {Kitaguchi}, {Koglin}, {Madsen},
  {Mao}, {Miyasaka}, {Mori}, {Perri}, {Pivovaroff}, {Puccetti}, {Rana},
  {Stern}, {Tagliaferri}, {Westergaard}, {Zhang}, {Zoglauer}, {Gurwell},
  {Uemura}, {Akitaya}, {Kawabata}, {Kawaguchi}, {Kanda}, {Moritani}, {Takaki},
  {Ui}, {Yoshida}, {Agarwal}, \& {Gupta}}]{2015ApJ...807...79H}
{Hayashida}, M., {Nalewajko}, K., {Madejski}, G.~M., {et~al.} 2015, \apj, 807,
  79, \dodoi{10.1088/0004-637X/807/1/79}

\bibitem[{{H.E.S.S. Collaboration} {et~al.}(2019){H.E.S.S. Collaboration},
  {Abdalla}, {Adam}, {Aharonian}, {Ait Benkhali}, {Ang{\"u}ner}, {Arakawa},
  {Arcaro}, {Armand}, {Ashkar}, {Backes}, {Barbosa Martins}, {Barnard},
  {Becherini}, {Berge}, {Bernl{\"o}hr}, {Blackwell}, {B{\"o}ttcher}, {Boisson},
  {Bolmont}, {Bonnefoy}, {Bregeon}, {Breuhaus}, {Brun}, {Brun}, {Bryan},
  {B{\"u}chele}, {Bulik}, {Bylund}, {Capasso}, {Caroff}, {Carosi}, {Casanova},
  {Cerruti}, {Chand}, {Chandra}, {Chen}, {Colafrancesco}, {Cury{\l}o},
  {Davids}, {Deil}, {Devin}, {deWilt}, {Dirson}, {Djannati-Ata{\"\i}},
  {Dmytriiev}, {Donath}, {Doroshenko}, {Drury}, {Dyks}, {Egberts}, {Emery},
  {Ernenwein}, {Eschbach}, {Feijen}, {Fegan}, {Fiasson}, {Fontaine}, {Funk},
  {F{\"u}{\ss}ling}, {Gabici}, {Gallant}, {Gat{\'e}}, {Giavitto}, {Glawion},
  {Glicenstein}, {Gottschall}, {Grondin}, {Hahn}, {Haupt}, {Heinzelmann},
  {Henri}, {Hermann}, {Hinton}, {Hofmann}, {Hoischen}, {Holch}, {Holler},
  {Horns}, {Huber}, {Iwasaki}, {Jamrozy}, {Jankowsky}, {Jankowsky},
  {Jardin-Blicq}, {Jung-Richardt}, {Kastendieck}, {Katarzy{\'n}ski},
  {Katsuragawa}, {Katz}, {Khangulyan}, {Kh{\'e}lifi}, {King}, {Klepser},
  {Klu{\'z}niak}, {Komin}, {Kosack}, {Kostunin}, {Kraus}, {Lamanna}, {Lau},
  {Lemi{\`e}re}, {Lemoine-Goumard}, {Lenain}, {Leser}, {Levy}, {Lohse},
  {Lypova}, {Mackey}, {Majumdar}, {Malyshev}, {Marandon}, {Marcowith}, {Mares},
  {Mariaud}, {Mart{\'\i}-Devesa}, {Marx}, {Maurin}, {Meintjes}, {Mitchell},
  {Moderski}, {Mohamed}, {Mohrmann}, {Moore}, {Moulin}, {Muller}, {Murach},
  {Nakashima}, {de Naurois}, {Ndiyavala}, {Niederwanger}, {Niemiec}, {Oakes},
  {O'Brien}, {Odaka}, {Ohm}, {de Ona Wilhelmi}, {Ostrowski}, {Oya}, {Panter},
  {Parsons}, {Perennes}, {Petrucci}, {Peyaud}, {Piel}, {Pita}, {Poireau},
  {Priyana Noel}, {Prokhorov}, {Prokoph}, {P{\"u}hlhofer}, {Punch},
  {Quirrenbach}, {Raab}, {Rauth}, {Reimer}, {Reimer}, {Remy}, {Renaud},
  {Rieger}, {Rinchiuso}, {Romoli}, {Rowell}, {Rudak}, {Ruiz-Velasco},
  {Sahakian}, {Saito}, {Sanchez}, {Santangelo}, {Sasaki}, {Schlickeiser},
  {Sch{\"u}ssler}, {Schulz}, {Schutte}, {Schwanke}, {Schwemmer},
  {Seglar-Arroyo}, {Senniappan}, {Seyffert}, {Shafi}, {Shiningayamwe},
  {Simoni}, {Sinha}, {Sol}, {Specovius}, {Spir-Jacob}, {Stawarz}, {Steenkamp},
  {Stegmann}, {Steppa}, {Takahashi}, {Tavernier}, {Taylor}, {Terrier},
  {Tiziani}, {Tluczykont}, {Trichard}, {Tsirou}, {Tsuji}, {Tuffs}, {Uchiyama},
  {van der Walt}, {van Eldik}, {van Rensburg}, {van Soelen}, {Vasileiadis},
  {Veh}, {Venter}, {Vincent}, {Vink}, {Voisin}, {V{\"o}lk}, {Vuillaume},
  {Wadiasingh}, {Wagner}, {White}, {Wierzcholska}, {Yang}, {Yoneda},
  {Zacharias}, {Zanin}, {Zdziarski}, {Zech}, {Ziegler}, {Zorn}, {{\.Z}ywucka},
  \& {Meyer}}]{2019A&A...627A.159H}
{H.E.S.S. Collaboration}, {Abdalla}, H., {Adam}, R., {et~al.} 2019, \aap, 627,
  A159, \dodoi{10.1051/0004-6361/201935704}

\bibitem[{{Hirotani} {et~al.}(2021){Hirotani}, {Krasnopolsky}, {Shang},
  {Nishikawa}, \& {Watson}}]{2021ApJ...908...88H}
{Hirotani}, K., {Krasnopolsky}, R., {Shang}, H., {Nishikawa}, K.-i., \&
  {Watson}, M. 2021, \apj, 908, 88, \dodoi{10.3847/1538-4357/abd3a6}

\bibitem[{{Hunter}(2007)}]{2007CSE.....9...90H}
{Hunter}, J.~D. 2007, Computing in Science and Engineering, 9, 90,
  \dodoi{10.1109/MCSE.2007.55}

\bibitem[{{Kaspi} {et~al.}(2007){Kaspi}, {Brandt}, {Maoz}, {Netzer},
  {Schneider}, \& {Shemmer}}]{2007ApJ...659..997K}
{Kaspi}, S., {Brandt}, W.~N., {Maoz}, D., {et~al.} 2007, \apj, 659, 997,
  \dodoi{10.1086/512094}

\bibitem[{{Kilerci Eser} {et~al.}(2015){Kilerci Eser}, {Vestergaard},
  {Peterson}, {Denney}, \& {Bentz}}]{2015ApJ...801....8K}
{Kilerci Eser}, E., {Vestergaard}, M., {Peterson}, B.~M., {Denney}, K.~D., \&
  {Bentz}, M.~C. 2015, \apj, 801, 8, \dodoi{10.1088/0004-637X/801/1/8}

\bibitem[{{Kim} {et~al.}(2020){Kim}, {Krichbaum}, {Broderick}, {Wielgus},
  {Blackburn}, {G{\'o}mez}, {Johnson}, {Bouman}, {Chael}, {Akiyama}, {Jorstad},
  {Marscher}, {Issaoun}, {Janssen}, {Chan}, {Savolainen}, {Pesce}, {{\"O}zel},
  {Alberdi}, {Alef}, {Asada}, {Azulay}, {Baczko}, {Ball}, {Balokovi{\'c}},
  {Barrett}, {Bintley}, {Boland }, {Bower}, {Bremer}, {Brinkerink},
  {Brissenden}, {Britzen}, {Broguiere}, {Bronzwaer}, {Byun}, {Carlstrom},
  {Chatterjee}, {Chatterjee}, {Chen}, {Chen}, {Cho}, {Christian}, {Conway},
  {Cordes}, {Crew}, {Cui}, {Davelaar}, {De Laurentis}, {Deane}, {Dempsey},
  {Desvignes}, {Dexter}, {Doeleman}, {Eatough}, {Falcke}, {Fish}, {Fomalont},
  {Fraga-Encinas}, {Friberg}, {Fromm}, {Galison}, {Gammie}, {Garc{\'\i}a},
  {Gentaz}, {Georgiev}, {Goddi}, {Gold}, {G{\'o}mez-Ruiz}, {Gu}, {Gurwell},
  {Hada}, {Hecht}, {Hesper}, {Ho}, {Ho}, {Honma}, {Huang}, {Huang}, {Hughes},
  {Ikeda}, {Inoue}, {James}, {Jannuzi}, {Jeter}, {Jiang}, {Jimenez-Rosales},
  {Jung}, {Karami}, {Karuppusamy}, {Kawashima}, {Keating}, {Kettenis}, {Kim},
  {Kim}, {Kino}, {Koay}, {Koch}, {Koyama}, {Kramer}, {Kramer}, {Kuo}, {Lauer},
  {Lee}, {Li}, {Li}, {Lindqvist}, {Lico}, {Liu}, {Liuzzo}, {Lo}, {Lobanov},
  {Loinard}, {Lonsdale}, {Lu}, {MacDonald}, {Mao}, {Markoff}, {Marrone},
  {Mart{\'\i}-Vidal}, {Matsushita}, {Matthews}, {Medeiros}, {Menten}, {Mizuno},
  {Mizuno}, {Moran}, {Moriyama}, {Moscibrodzka}, {Musoke}, {M{\"u}ller},
  {Nagai}, {Nagar}, {Nakamura}, {Narayan}, {Narayanan}, {Natarajan}, {Neri},
  {Ni}, {Noutsos}, {Okino}, {Olivares}, {Ortiz-Le{\'o}n}, {Oyama}, {Palumbo},
  {Park}, {Patel}, {Pen}, {Pi{\'e}tu}, {Plambeck}, {PopStefanija}, {Porth},
  {Prather}, {Preciado-L{\'o}pez}, {Psaltis}, {Pu}, {Ramakrishnan}, {Rao},
  {Rawlings}, {Raymond}, {Rezzolla}, {Ripperda}, {Roelofs}, {Rogers}, {Ros},
  {Rose}, {Roshanineshat}, {Rottmann}, {Roy}, {Ruszczyk}, {Ryan}, {Rygl},
  {S{\'a}nchez}, {S{\'a}nchez-Arguelles}, {Sasada}, {Schloerb}, {Schuster},
  {Shao}, {Shen}, {Small}, {Sohn}, {SooHoo}, {Tazaki}, {Tiede}, {Tilanus},
  {Titus}, {Toma}, {Torne}, {Trent}, {Traianou}, {Trippe}, {Tsuda}, {van
  Bemmel}, {van Langevelde}, {van Rossum}, {Wagner}, {Wardle}, {Ward-Thompson},
  {Weintroub}, {Wex}, {Wharton}, {Wong}, {Wu}, {Yoon}, {Young}, {Young},
  {Younsi}, {Yuan}, {Yuan}, {Zensus}, {Zhao}, {Zhao}, {Zhu}, {Algaba},
  {Allardi}, {Amestica}, {Anczarski}, {Bach}, {Baganoff}, {Beaudoin}, {Benson},
  {Berthold}, {Blanchard}, {Blundell}, {Bustamente}, {Cappallo},
  {Castillo-Dom{\'\i}nguez}, {Chang}, {Chang}, {Chang}, {Chen}, {Chilson},
  {Chuter}, {Rosado}, {Coulson}, {Crowley}, {Derome}, {Dexter}, {Dornbusch},
  {Dudevoir}, {Dzib}, {Eckart}, {Eckert}, {Erickson}, {Everett}, {Faber},
  {Farah}, {Fath}, {Folkers}, {Forbes}, {Freund}, {Gale}, {Gao}, {Geertsema},
  {Graham}, {Greer}, {Grosslein}, {Gueth}, {Haggard}, {Halverson}, {Han},
  {Han}, {Hao}, {Hasegawa}, {Henning}, {Hern{\'a}ndez-G{\'o}mez},
  {Herrero-Illana}, {Heyminck}, {Hirota}, {Hoge}, {Huang}, {Violette
  Impellizzeri}, {Jiang}, {John}, {Kamble}, {Keisler}, {Kimura}, {Kono},
  {Kubo}, {Kuroda}, {Lacasse}, {Laing}, {Leitch}, {Li}, {Lin}, {Liu}, {Liu},
  {Lu}, {Marson}, {Martin-Cocher}, {Massingill}, {Matulonis}, {McColl},
  {McWhirter}, {Messias}, {Meyer-Zhao}, {Michalik}, {Monta{\~n}a},
  {Montgomerie}, {Mora-Klein}, {Muders}, {Nadolski}, {Navarro}, {Neilsen},
  {Nguyen}, {Nishioka}, {Norton}, {Nowak}, {Nystrom}, {Ogawa}, {Oshiro},
  {Oyama}, {Parsons}, {Pe{\~n}alver}, {Phillips}, {Poirier}, {Pradel},
  {Primiani}, {Raffin}, {Rahlin}, {Reiland}, {Risacher}, {Ruiz},
  {S{\'a}ez-Mada{\'\i}n}, {Sassella}, {Schellart}, {Shaw}, {Silva}, {Shiokawa},
  {Smith}, {Snow}, {Souccar}, {Sousa}, {Sridharan}, {Srinivasan}, {Stahm},
  {Stark}, {Story}, {Timmer}, {Vertatschitsch}, {Walther}, {Wei}, {Whitehorn},
  {Whitney}, {Woody}, {Wouterloot}, {Wright}, {Yamaguchi}, {Yu}, {Zeballos},
  {Zhang}, {Ziurys}, \& {Event Horizon Telescope
  Collaboration}}]{2020A&A...640A..69K}
{Kim}, J.-Y., {Krichbaum}, T.~P., {Broderick}, A.~E., {et~al.} 2020, \aap, 640,
  A69, \dodoi{10.1051/0004-6361/202037493}

\bibitem[{{Landi} {et~al.}(2012){Landi}, {Del Zanna}, {Young}, {Dere}, \&
  {Mason}}]{2012ApJ...744...99L}
{Landi}, E., {Del Zanna}, G., {Young}, P.~R., {Dere}, K.~P., \& {Mason}, H.~E.
  2012, \apj, 744, 99, \dodoi{10.1088/0004-637X/744/2/99}

\bibitem[{{Lefa} {et~al.}(2011){Lefa}, {Rieger}, \&
  {Aharonian}}]{2011ApJ...740...64L}
{Lefa}, E., {Rieger}, F.~M., \& {Aharonian}, F. 2011, \apj, 740, 64,
  \dodoi{10.1088/0004-637X/740/2/64}

\bibitem[{{Lei} \& {Wang}(2015)}]{2015PASJ...67...79L}
{Lei}, M., \& {Wang}, J. 2015, \pasj, 67, 79, \dodoi{10.1093/pasj/psv055}

\bibitem[{{Liu} \& {Bai}(2006)}]{2006ApJ...653.1089L}
{Liu}, H.~T., \& {Bai}, J.~M. 2006, \apj, 653, 1089, \dodoi{10.1086/509097}

\bibitem[{{Lovelace}(1976)}]{1976Natur.262..649L}
{Lovelace}, R.~V.~E. 1976, \nat, 262, 649, \dodoi{10.1038/262649a0}

\bibitem[{{MAGIC Collaboration} {et~al.}(2008){MAGIC Collaboration}, {Albert},
  {Aliu}, {Anderhub}, {Antonelli}, {Antoranz}, {Backes}, {Baixeras}, {Barrio},
  {Bartko}, {Bastieri}, {Becker}, {Bednarek}, {Berger}, {Bernardini},
  {Bigongiari}, {Biland}, {Bock}, {Bonnoli}, {Bordas}, {Bosch-Ramon}, {Bretz},
  {Britvitch}, {Camara}, {Carmona}, {Chilingarian}, {Commichau}, {Contreras},
  {Cortina}, {Costado}, {Covino}, {Curtef}, {Dazzi}, {De Angelis}, {de Cea del
  Pozo}, {de los Reyes}, {De Lotto}, {De Maria}, {De Sabata}, {Delgado Mendez},
  {Dominguez}, {Dorner}, {Doro}, {Errando}, {Fagiolini}, {Ferenc},
  {Fern{\'a}ndez}, {Firpo}, {Fonseca}, {Font}, {Galante}, {Garc{\'\i}a
  L{\'o}pez}, {Garczarczyk}, {Gaug}, {Goebel}, {Hayashida}, {Herrero},
  {H{\"o}hne}, {Hose}, {Hsu}, {Huber}, {Jogler}, {Kneiske}, {Kranich}, {La
  Barbera}, {Laille}, {Leonardo}, {Lindfors}, {Lombardi}, {Longo}, {L{\'o}pez},
  {Lorenz}, {Majumdar}, {Maneva}, {Mankuzhiyil}, {Mannheim}, {Maraschi},
  {Mariotti}, {Mart{\'\i}nez}, {Mazin}, {Meucci}, {Meyer}, {Miranda},
  {Mirzoyan}, {Mizobuchi}, {Moles}, {Moralejo}, {Nieto}, {Nilsson}, {Ninkovic},
  {Otte}, {Oya}, {Panniello}, {Paoletti}, {Paredes}, {Pasanen}, {Pascoli},
  {Pauss}, {Pegna}, {Perez-Torres}, {Persic}, {Peruzzo}, {Piccioli}, {Prada},
  {Prandini}, {Puchades}, {Raymers}, {Rhode}, {Rib{\'o}}, {Rico}, {Rissi},
  {Robert}, {R{\"u}gamer}, {Saggion}, {Saito}, {Salvati}, {Sanchez-Conde},
  {Sartori}, {Satalecka}, {Scalzotto}, {Scapin}, {Schmitt}, {Schweizer},
  {Shayduk}, {Shinozaki}, {Shore}, {Sidro}, {Sierpowska-Bartosik},
  {Sillanp{\"a}{\"a}}, {Sobczynska}, {Spanier}, {Stamerra}, {Stark}, {Takalo},
  {Tavecchio}, {Temnikov}, {Tescaro}, {Teshima}, {Tluczykont}, {Torres},
  {Turini}, {Vankov}, {Venturini}, {Vitale}, {Wagner}, {Wittek}, {Zabalza},
  {Zandanel}, {Zanin}, \& {Zapatero}}]{2008Sci...320.1752M}
{MAGIC Collaboration}, {Albert}, J., {Aliu}, E., {et~al.} 2008, Science, 320,
  1752, \dodoi{10.1126/science.1157087}

\bibitem[{{Mannheim}(1993)}]{1993A&A...269...67M}
{Mannheim}, K. 1993, \aap, 269, 67.
\newblock \doarXiv{astro-ph/9302006}

\bibitem[{{Marcowith} {et~al.}(1995){Marcowith}, {Henri}, \&
  {Pelletier}}]{1995MNRAS.277..681M}
{Marcowith}, A., {Henri}, G., \& {Pelletier}, G. 1995, \mnras, 277, 681,
  \dodoi{10.1093/mnras/277.2.681}

\bibitem[{{Marscher}(2014)}]{2014ApJ...780...87M}
{Marscher}, A.~P. 2014, \apj, 780, 87, \dodoi{10.1088/0004-637X/780/1/87}

\bibitem[{{Marziani} {et~al.}(1996){Marziani}, {Sulentic}, {Dultzin-Hacyan},
  {Calvani}, \& {Moles}}]{1996ApJS..104...37M}
{Marziani}, P., {Sulentic}, J.~W., {Dultzin-Hacyan}, D., {Calvani}, M., \&
  {Moles}, M. 1996, \apjs, 104, 37, \dodoi{10.1086/192291}

\bibitem[{{Mastichiadis} \& {Kirk}(1995)}]{1995A&A...295..613M}
{Mastichiadis}, A., \& {Kirk}, J.~G. 1995, \aap, 295, 613.
\newblock \url{https://ui.adsabs.harvard.edu/abs/1995A&A...295..613M}

\bibitem[{{Matthews} {et~al.}(2020){Matthews}, {Bell}, \&
  {Blundell}}]{2020NewAR..8901543M}
{Matthews}, J.~H., {Bell}, A.~R., \& {Blundell}, K.~M. 2020, \nar, 89, 101543,
  \dodoi{10.1016/j.newar.2020.101543}

\bibitem[{{Mattox} {et~al.}(1996){Mattox}, {Bertsch}, {Chiang}, {Dingus},
  {Digel}, {Esposito}, {Fierro}, {Hartman}, {Hunter}, {Kanbach}, {Kniffen},
  {Lin}, {Macomb}, {Mayer-Hasselwander}, {Michelson}, {von Montigny},
  {Mukherjee}, {Nolan}, {Ramanamurthy}, {Schneid}, {Sreekumar}, {Thompson}, \&
  {Willis}}]{1996ApJ...461..396M}
{Mattox}, J.~R., {Bertsch}, D.~L., {Chiang}, J., {et~al.} 1996, \apj, 461, 396,
  \dodoi{10.1086/177068}

\bibitem[{{Meyer} {et~al.}(2021){Meyer}, {Petropoulou}, \&
  {Christie}}]{2021ApJ...912...40M}
{Meyer}, M., {Petropoulou}, M., \& {Christie}, I.~M. 2021, \apj, 912, 40,
  \dodoi{10.3847/1538-4357/abedab}

\bibitem[{{Meyer} {et~al.}(2019){Meyer}, {Scargle}, \&
  {Blandford}}]{2019ApJ...877...39M}
{Meyer}, M., {Scargle}, J.~D., \& {Blandford}, R.~D. 2019, \apj, 877, 39,
  \dodoi{10.3847/1538-4357/ab1651}

\bibitem[{Oliphant(2015)}]{10.5555/2886196}
Oliphant, T.~E. 2015, Guide to NumPy, 2nd edn. (North Charleston, SC, USA:
  CreateSpace Independent Publishing Platform)

\bibitem[{{Pati{\~n}o-{\'A}lvarez} {et~al.}(2019){Pati{\~n}o-{\'A}lvarez},
  {Dzib}, {Lobanov}, \& {Chavushyan}}]{2019A&A...630A..56P}
{Pati{\~n}o-{\'A}lvarez}, V.~M., {Dzib}, S.~A., {Lobanov}, A., \& {Chavushyan},
  V. 2019, \aap, 630, A56, \dodoi{10.1051/0004-6361/201834401}

\bibitem[{{Pian} {et~al.}(2005){Pian}, {Falomo}, \&
  {Treves}}]{2005MNRAS.361..919P}
{Pian}, E., {Falomo}, R., \& {Treves}, A. 2005, \mnras, 361, 919,
  \dodoi{10.1111/j.1365-2966.2005.09216.x}

\bibitem[{{Pittori} {et~al.}(2018){Pittori}, {Lucarelli}, {Verrecchia},
  {Raiteri}, {Villata}, {Vittorini}, {Tavani}, {Puccetti}, {Perri},
  {Donnarumma}, {Vercellone}, {Acosta-Pulido}, {Bachev}, {Ben{\'\i}tez},
  {Borman}, {Carnerero}, {Carosati}, {Chen}, {Ehgamberdiev}, {Goded},
  {Grishina}, {Hiriart}, {Hsiao}, {Jorstad}, {Kimeridze}, {Kopatskaya},
  {Kurtanidze}, {Kurtanidze}, {Larionov}, {Larionova}, {Marscher},
  {Mirzaqulov}, {Morozova}, {Nilsson}, {Samal}, {Sigua}, {Spassov},
  {Strigachev}, {Takalo}, {Antonelli}, {Bulgarelli}, {Cattaneo},
  {Colafrancesco}, {Giommi}, {Longo}, {Morselli}, \&
  {Paoletti}}]{2018ApJ...856...99P}
{Pittori}, C., {Lucarelli}, F., {Verrecchia}, F., {et~al.} 2018, \apj, 856, 99,
  \dodoi{10.3847/1538-4357/aab1f9}

\bibitem[{{Poutanen} \& {Stern}(2010)}]{2010ApJ...717L.118P}
{Poutanen}, J., \& {Stern}, B. 2010, \apjl, 717, L118,
  \dodoi{10.1088/2041-8205/717/2/L118}

\bibitem[{{Rani} {et~al.}(2018){Rani}, {Jorstad}, {Marscher}, {Agudo},
  {Sokolovsky}, {Larionov}, {Smith}, {Mosunova}, {Borman}, {Grishina},
  {Kopatskaya}, {Mokrushina}, {Morozova}, {Savchenko}, {Troitskaya},
  {Troitsky}, {Thum}, {Molina}, \& {Casadio}}]{2018ApJ...858...80R}
{Rani}, B., {Jorstad}, S.~G., {Marscher}, A.~P., {et~al.} 2018, \apj, 858, 80,
  \dodoi{10.3847/1538-4357/aab785}

\bibitem[{{Reimer}(2007)}]{2007ApJ...665.1023R}
{Reimer}, A. 2007, \apj, 665, 1023, \dodoi{10.1086/519766}

\bibitem[{{Rieger}(2019)}]{2019Galax...7...78R}
{Rieger}, F.~M. 2019, Galaxies, 7, 78, \dodoi{10.3390/galaxies7030078}

\bibitem[{{Rieger} {et~al.}(2007){Rieger}, {Bosch-Ramon}, \&
  {Duffy}}]{2007Ap&SS.309..119R}
{Rieger}, F.~M., {Bosch-Ramon}, V., \& {Duffy}, P. 2007, \apss, 309, 119,
  \dodoi{10.1007/s10509-007-9466-z}

\bibitem[{{Rieger} \& {Mannheim}(2002)}]{2002A&A...396..833R}
{Rieger}, F.~M., \& {Mannheim}, K. 2002, \aap, 396, 833,
  \dodoi{10.1051/0004-6361:20021457}

\bibitem[{{Shah} {et~al.}(2019){Shah}, {Jithesh}, {Sahayanathan}, {Misra}, \&
  {Iqbal}}]{2019MNRAS.484.3168S}
{Shah}, Z., {Jithesh}, V., {Sahayanathan}, S., {Misra}, R., \& {Iqbal}, N.
  2019, \mnras, 484, 3168, \dodoi{10.1093/mnras/stz151}

\bibitem[{{Shukla} \& {Mannheim}(2020)}]{2020NatCo..11.4176S}
{Shukla}, A., \& {Mannheim}, K. 2020, Nature Communications, 11, 4176,
  \dodoi{10.1038/s41467-020-17912-z}

\bibitem[{{Stern} \& {Poutanen}(2011)}]{2011MNRAS.417L..11S}
{Stern}, B.~E., \& {Poutanen}, J. 2011, \mnras, 417, L11,
  \dodoi{10.1111/j.1745-3933.2011.01107.x}

\bibitem[{{Stern} \& {Poutanen}(2014)}]{2014ApJ...794....8S}
---. 2014, \apj, 794, 8, \dodoi{10.1088/0004-637X/794/1/8}

\bibitem[{{Summerlin} \& {Baring}(2012)}]{2012ApJ...745...63S}
{Summerlin}, E.~J., \& {Baring}, M.~G. 2012, \apj, 745, 63,
  \dodoi{10.1088/0004-637X/745/1/63}

\bibitem[{{Tan} {et~al.}(2020){Tan}, {Xue}, {Du}, {Xi}, {Wang}, \&
  {Xie}}]{2020ApJS..248...27T}
{Tan}, C., {Xue}, R., {Du}, L.-M., {et~al.} 2020, \apjs, 248, 27,
  \dodoi{10.3847/1538-4365/ab8cc6}

\bibitem[{{Tavecchio} \& {Ghisellini}(2008)}]{2008MNRAS.386..945T}
{Tavecchio}, F., \& {Ghisellini}, G. 2008, \mnras, 386, 945,
  \dodoi{10.1111/j.1365-2966.2008.13072.x}

\bibitem[{{Tavecchio} \& {Ghisellini}(2012)}]{2012arXiv1209.2291T}
---. 2012, arXiv e-prints, arXiv:1209.2291.
\newblock \doarXiv{1209.2291}

\bibitem[{{Tavecchio} \& {Mazin}(2009)}]{2009MNRAS.392L..40T}
{Tavecchio}, F., \& {Mazin}, D. 2009, \mnras, 392, L40,
  \dodoi{10.1111/j.1745-3933.2008.00584.x}

\bibitem[{{Telfer} {et~al.}(2002){Telfer}, {Zheng}, {Kriss}, \&
  {Davidsen}}]{2002ApJ...565..773T}
{Telfer}, R.~C., {Zheng}, W., {Kriss}, G.~A., \& {Davidsen}, A.~F. 2002, \apj,
  565, 773, \dodoi{10.1086/324689}

\bibitem[{{Vuillaume} {et~al.}(2018){Vuillaume}, {Henri}, \&
  {Petrucci}}]{2018A&A...620A..41V}
{Vuillaume}, T., {Henri}, G., \& {Petrucci}, P.~O. 2018, \aap, 620, A41,
  \dodoi{10.1051/0004-6361/201731899}

\bibitem[{{Wendel} {et~al.}(2021){Wendel}, {Becerra Gonz{\'a}lez}, {Paneque},
  \& {Mannheim}}]{2021A&A...646A.115W}
{Wendel}, C., {Becerra Gonz{\'a}lez}, J., {Paneque}, D., \& {Mannheim}, K.
  2021, \aap, 646, A115, \dodoi{10.1051/0004-6361/202038343}

\bibitem[{{Zdziarski}(1988)}]{1988ApJ...335..786Z}
{Zdziarski}, A.~A. 1988, \apj, 335, 786, \dodoi{10.1086/166967}

\end{thebibliography}

%\listofchanges

\end{document}